\documentclass[fleqn,usenatbib]{mnras}

\usepackage{newtxtext,newtxmath}

\usepackage[T1]{fontenc}

\DeclareRobustCommand{\VAN}[3]{#2}
\let\VANthebibliography\thebibliography
\def\thebibliography{\DeclareRobustCommand{\VAN}[3]{##3}\VANthebibliography}

\usepackage{graphicx}	
\usepackage{amsmath}	

\title[New tools for generating and modelling metal lines]{Towards an automatic approach to modelling the circumgalactic medium: new tools for mock making and fitting of metal profiles in large surveys}

\author[A. L. Longobardi et al.]{
Alessia Longobardi,$^{1,2}$\thanks{E-mail: alessia.longobardi@unimib.it}
Matteo Fossati,$^{1,2}$
Michele Fumagalli,$^{1,3}$ 
Bhaskar Agarwal,$^{4}$
Emma Lofthouse,$^{1,2}$ \and
Marta Galbiati,$^{1}$
Rajeshwari Dutta,$^{1,2}$ 
Trystyn A. M. Berg, $^{1,2}$
Louise A. Welsh $^{1,2}$
\\
$^{1}$Dipartimento di Fisica G. Occhialini, Universit\`a degli Studi di Milano-Bicocca, Piazza della Scienza 3, 20126 Milano, Italy \\
$^{2}$INAF - Osservatorio Astronomico di Brera, via Bianchi 46, 23087 Merate (LC), Italy \\
$^{3}$INAF - Osservatorio Astronomico di Trieste, via G. B. Tiepolo 11, 34143 Trieste, Italy \\
$^{4}$ Cineca, Via Magnanelli, 6/3, 40033 Casalecchio di Reno BO, Italy\\
}

\date{Accepted XXX. Received YYY; in original form ZZZ}

\pubyear{\the\year{}}

\begin{document}
\label{firstpage}
\pagerange{\pageref{firstpage}--\pageref{lastpage}}
\maketitle

\begin{abstract}
We present two new tools for studying and modelling metal absorption lines in the circumgalactic medium. 
The first tool, dubbed ``NMF Profile Maker'' (NMF$-$PM), uses a non-negative matrix factorization (NMF) method and provides a robust means to generate large libraries of realistic metal absorption profiles. The method is trained and tested on 650 unsaturated metal absorbers in the redshift interval $z=0.9-4.2$ with column densities between $11.2 \le \log{(\mathrm{N/cm^{-2}})} \le 16.3$, obtained from high-resolution ($R> 4000$) and high signal-to-noise ratio ($S/N \ge 10$) quasar spectroscopy. To avoid spurious features, we train on infinite $S/N$ Voigt models of the observed line profiles derived using the code ``Monte-Carlo Absorption Line Fitter'' (MC$-$ALF), a novel automatic Bayesian fitting code that is the second tool we present in this work. MC$-$ALF is a Monte Carlo code based on nested sampling that, without the need for any prior guess or human intervention, can decompose metal lines into individual Voigt components.  Both MC$-$ALF and NMF$-$PM are made publicly available to allow the community to produce large libraries of synthetic metal profiles and to reconstruct Voigt models of absorption lines in an automatic fashion. Both tools contribute to the scientific effort of simulating and analysing metal absorbers in very large spectroscopic surveys of quasars like the ongoing Dark Energy Spectroscopic Instrument (DESI), the 4-meter
Multi-Object Spectroscopic Telescope (4MOST), and the WHT Enhanced Area Velocity Explorer (WEAVE) surveys. 

\end{abstract}




\section{Introduction}
In the current Cold Dark Matter (CDM) paradigm of structures' formation and evolution, the modelling of baryons is representing a challenge due to the large array of physical processes affecting this baryonic component \citep[e.g.,][and references therein]{vogelsberger20}. As these processes (e.g., gas cooling, star formation, stellar/AGN feedback, and their interplay with gravity) combine to shape the morphological and physical properties of the galaxies as we observe them today, the detailed study of the gas environment has become a priority in the field of galaxies' evolution. 

The circumgalactic medium (CGM), i.e. the baryonic component that connects the galaxies' interstellar medium (ISM) with their large-scale environment (intergalactic medium, or IGM), plays a key role in our understanding of how galaxies evolve, by allowing us to follow the 'baryon cycle' in which gas cycles into, out of, and through galaxies. Observations of this component both in emission and absorption allow us to trace the thermal and chemical evolution of diffuse gas in the Universe as a function of redshift \citep[e.g.,][and references therein]{tumlinson17}. In the last two decades, an increasing effort has been put into the study of the CGM physical properties (temperature, density, metallicity) through the analysis of absorption lines in spectra of background sources like bright quasars, such as intervening metal absorbers, broad absorption line (BAL) quasars or \ion{H}{I} selected systems like damped Ly$\alpha$ absorbers (DLAs) or Lyman Limit systems (LLSs). Studies across these different populations enable access to a wide range of gas column densities, i.e. $\log{(\mathrm{N/cm^{-2}})} \ge 11$ \citep[e.g.,][]{sargent89,schaye00,prochaska05,simcoe11,rafelski12,fumagalli16b,dodorico22}, which in turn trace varying degrees of overdensities. 

With the advent of the Sloan Digital Sky Survey \citep[SDSS;][]{york00} and the compilation of the SDSS quasar catalogues \citep[final release by][]{lyke20}, astronomers now have access to just under a million spectroscopically confirmed quasars to statistically assess the properties of absorption line systems and trace the distribution and physical properties of the gaseous component around galaxies across the Universe \citep[e.g.][]{noterdaeme08,prochaska10,lan14,garnett17,anand21,anand22}.  These statistical studies have further enabled follow-up observations with high-resolution spectrographs on 8-and-10-m class telescopes that opened the possibility of identifying characteristic features in absorption systems while simultaneously allowing us to map the correlations between the galaxies and the ambient gaseous halos \citep[e.g.][]{werk16,prochaska17,fossati19,mackenzie19,rudie19,lofthouse20,lofthouse23,dutta20,wilde21}.

A boost to these studies is imminent thanks to large surveys at 4~-m class telescopes, like the Dark Energy Spectroscopic Instrument \citep[DESI;][]{DESI16} survey, the 4-meter Multi-Object Spectroscopic Telescope \citep[4MOST;][]{dejong12} survey, and the WHT Enhanced Area Velocity Explorer \citep[WEAVE;][]{dalton12,jin22} survey.
As the size and quality of the data increase, novel, fast, and efficient approaches for identifying and characterising absorption lines are needed. This is why an increasing effort has been put into proposing efficient approaches to: i) identify the different classes of quasars absorbers like metals \citep{cooksey13,zhu13a,zou21}, BALs \citep{guo19}, DLAs \citep{garnett17}, and LLSs \citep{fumagalli20} and ii) to derive models of the different line profiles that can accurately reproduce their behaviour in terms of maximum likelihood \citep[e.g.,][]{carswell14} or bayesian estimates \citep[][]{liang17}. Similarly, as samples increase, systematic errors overcome by far statistical uncertainties \citep[e.g.,][]{fumagalli20}. Tools are therefore required to create high-quality mocks to train and validate the science pipelines used to analyze the spectra and extract physical information on the absorption lines.

The objective of this paper is to contribute to the tools available in the field to tackle these challenges. Specifically, we present a new code, MC$-$ALF, which provides an automatic reconstruction of the shape of line profiles and extracts posterior distributions of the relevant physical parameters (such as the number of components, column density, and Doppler parameter of each component). Moreover, we introduce a new method, called NMF$-$PM, that generates synthetic but realistic-looking line profiles following a given distribution of desired line widths. Combined, these new codes provide useful pre- and post-processing tools to aid the science exploitation of future wide-field surveys, contributing both to the simulation of quasar spectra with realistic absorption lines, and to the higher-level analysis of downstream data products. In particular, our codes are tailored to the electronic transition lines of metal species (both low- and moderate-ion transitions, with ionization potential $\mathrm{IP} \lesssim 30-40\, eV$ and $\mathrm{IP} \sim 40-100\, eV$, respectively) that can be used to study the multi-phase nature of the CGM.


Our tools rely on advanced numerical techniques. Specifically, for MC$-$ALF, we adopt a Bayesian nested sampling approach to the line profile fitting. This method efficiently explores the full parameter space by slicing it into sub-volumes and fitting nested N$^{th}$ dimensional contours to identify the regions with a strong likelihood gradient where accurate sampling is required. 
For NMF$-$PM, instead, we rely on the Non-negative Matrix Factorization (NMF), which is a subclass of multivariate analysis techniques often associated with pattern recognition and blind source separation \citep{lee00}, also used within the astronomical community \citep[e.g.,][]{zhu13a,hurley14,ren18}. Offering us a well
established statistical framework for carrying out the representation
of positive and continuous signals, the NMF is an ideal algorithm for summarising the information contained in a large data set of metal absorbers and for carrying out robust modelling and prediction making. 

One of the largest difficulties in the development of synthetic metals' profiles via machine learning (ML) methods is to have a sample of training data, which satisfies both quality and quantity requirements. However, as part of three large and complete galaxies' surveys in quasar fields - the MUSE Analysis of Gas around Galaxies \citep[MAGG;][]{lofthouse20,dutta20} survey, the Quasar Sightline and Galaxy Evolution \citep[QSAGE;][]{bielby19,dutta21} survey, and the MUSE Ultra Deep Field \citep[MUDF;][]{fossati19} - our team has assembled a library of $\sim 700$ metal absorption lines representative of moderately to highly overdense gas. Thanks to this dataset, we are now able to train algorithms to generate metal profiles in quasar spectra with ML across a large variety of column densities, redshifts, and line widths.

The paper is structured as follows: in Section~\ref{data} we describe the spectroscopic surveys that provided the data to compile our library of metal systems. In Section~\ref{Voigt_fitting} we present and test the Voigt fitting algorithms at the basis of MC$-$ALF, which we used to produce a set of unsaturated metals profiles with infinite signal-to-noise ratio ($S/N$). In Section \ref{NMF_formalism}, we introduce instead NMF$-$PM,
presenting the NMF formulation useful to produce synthetic metal profiles. The latter are presented in Section \ref{NMF_results} and are discussed in the context of the upcoming large surveys of background quasars. A summary is presented in Section \ref{conclusions}.

\section{Library of absorption line systems}\label{data}

\begin{figure*}
    \includegraphics[trim={0.5cm 0cm 0.5cm 0cm},clip,width=\textwidth]{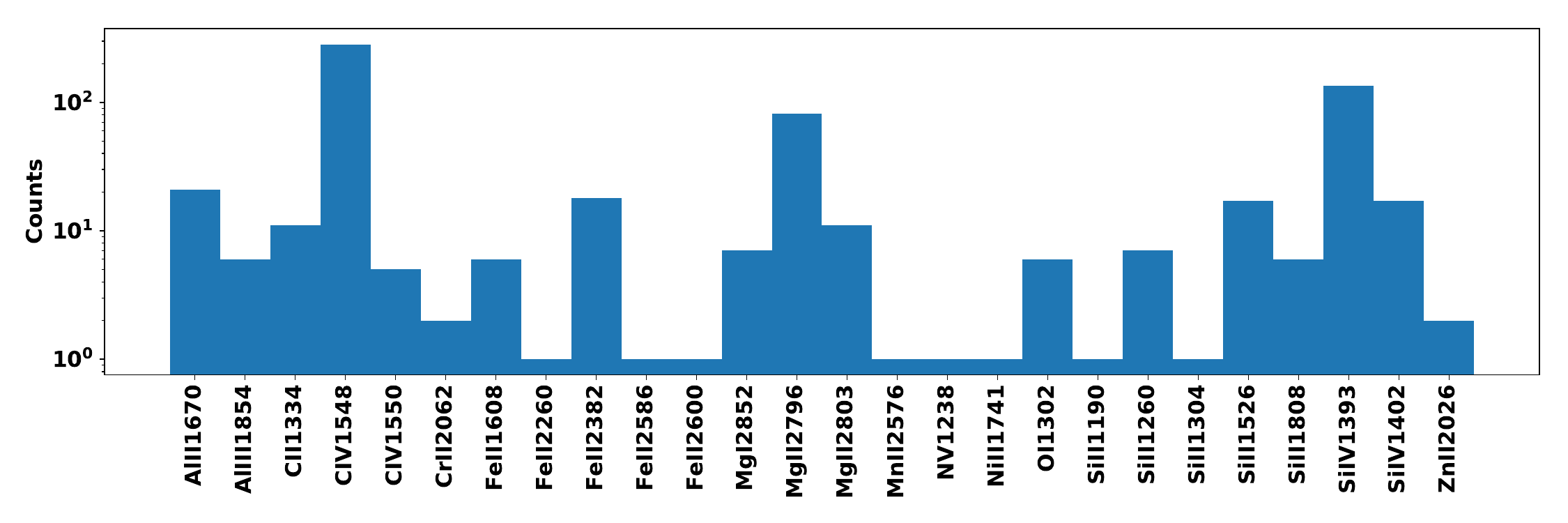}\\
	\includegraphics[trim={6.5cm 7.5cm 6.5cm 9cm},clip,width=\textwidth]{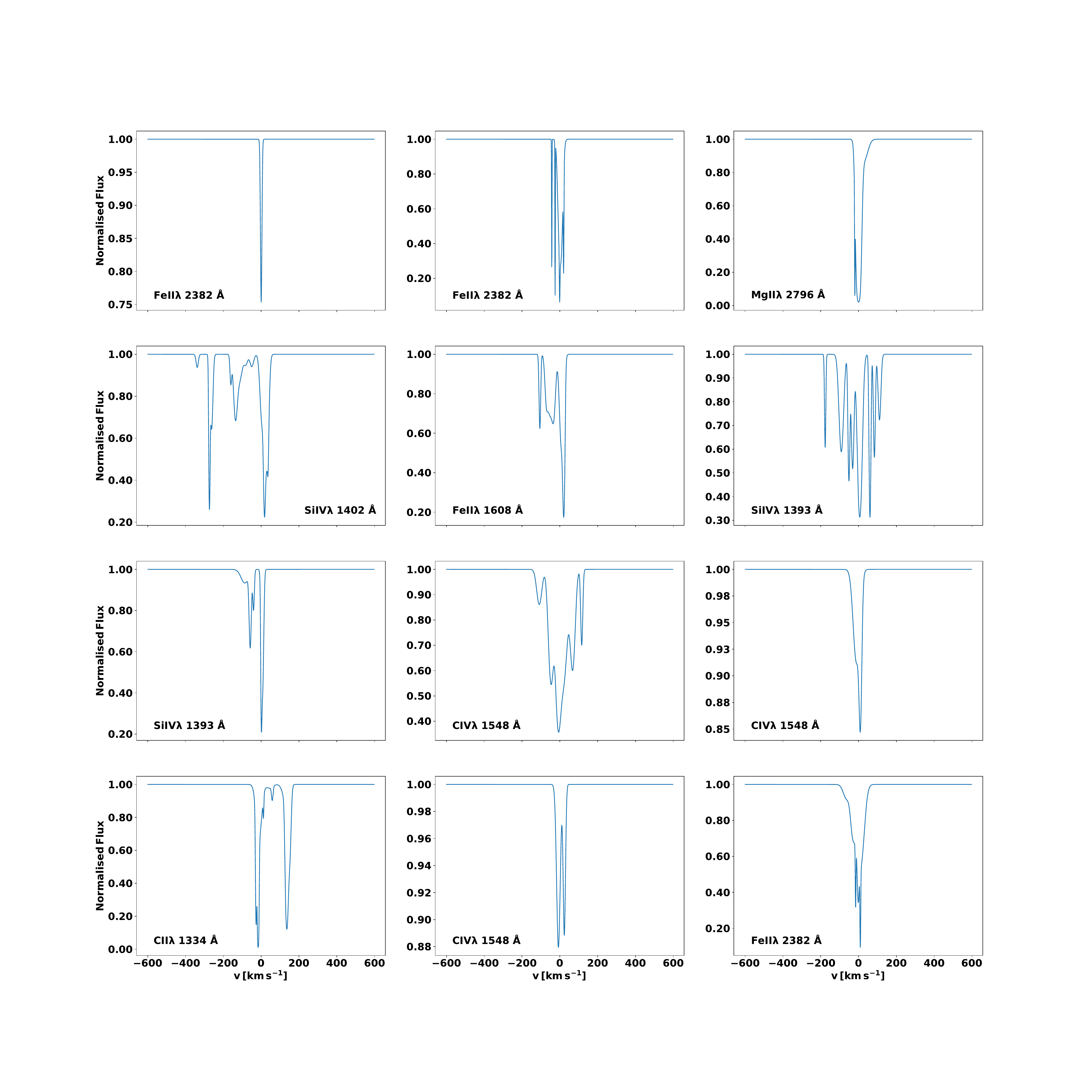}\\

    \caption{{\it Top Panel}: 650 absorbers in our sample: $\sim 70\%$ are moderate-ions, the remaining $\sim 30\%$ is represented by low-ions. {\it Bottom Panels}: Sub-sample of data fitted with MC$-$ALF. The sample gathers a large variety of profiles in terms of the number of components in each profile and $\mathrm{\Delta V_{90}}$ distribution.}
    \label{Sample}
\end{figure*}

\begin{figure*}
	\includegraphics[trim={6.5cm 1cm 6.5cm 2cm},clip,width=\textwidth]{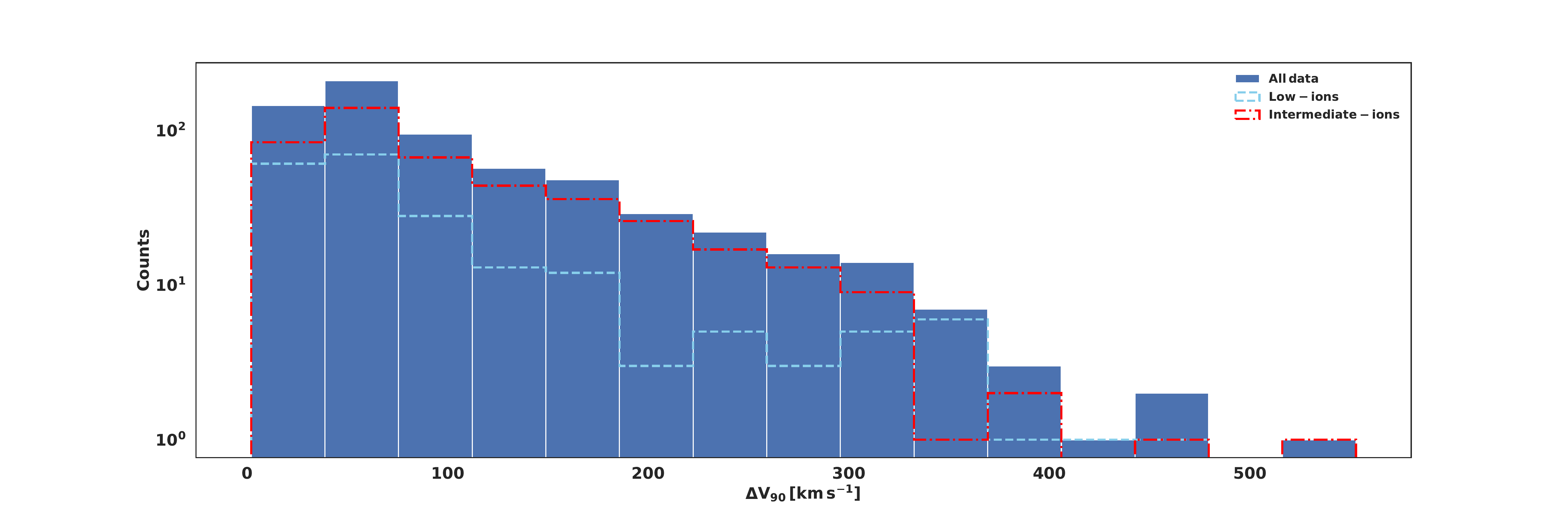}\\
	\includegraphics[trim={6.5cm 1cm 6.5cm 2cm},clip,width=\textwidth]{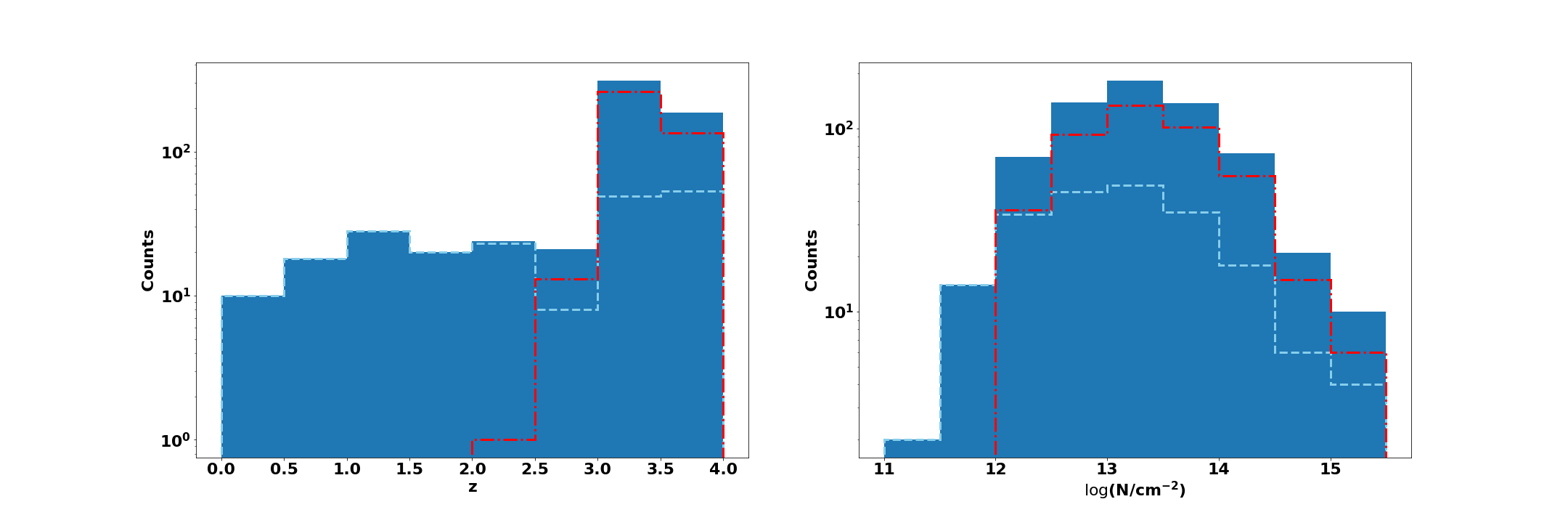}\\

    \caption{{\it Top Panel}: Distribution of  $\Delta\, \mathrm{V_{90}}$ values in our sample (dark blue). Dashed cyan and dotted-dashed red histograms show the $\Delta\, \mathrm{V_{90}}$ values for the low- and moderate-ions, respectively. {\it Bottom Panels}: Same as top panel, however, this time the histograms are relative to the redshift (left) and column density (right) distributions.}
    \label{Sample_dist}
\end{figure*}

\subsection{Spectroscopic surveys adopted}

For the purpose of developing, training, and testing our codes, we assemble a library of moderate-to-high $S/N$ spectra of absorption lines in different ionization stages and at different redshifts. Next, we provide a brief description of the compilation of the spectroscopic campaigns that form the dataset used in this work. We refer the reader to the listed references for additional details on data quality, and reduction techniques.

\subsubsection{The MAGG survey}

The MUSE Analysis of Gas around Galaxies (MAGG) survey \citep{lofthouse20} is based upon a MUSE Large Programme (ID 197.A-0384; PI Fumagalli) of 28 quasar fields at redshift $3.2 \le z \le 4.5$ for which $S/N \ge 10$ and medium- ($4000-10000$) or high-resolution ($20000-50000$) spectroscopy is available. High-resolution spectroscopy is a compilation of data from the Ultraviolet and Visual Echelle Spectrograph \citep[UVES;][]{dekker00}, the High-Resolution Echelle Spectrometer \citep[HIRES;][]{vogt94} and the Magellan Inamori Kyocera Echelle instruments \citep[MIKE;][]{Bernstein03}, while moderate resolution spectroscopy is from ESI \citep{sheinis02} and X-SHOOTER \citep{spano06,vernet11}. A total of  62 individual spectra were assembled for the 28 quasars.

Instrument-specific pipelines were used to carry out the data reduction, which included bias subtraction, flat- fielding, dark
subtraction (where applicable), and wavelength calibration. Once one-dimensional spectra were extracted, and eventually combined if multiple exposure were present, the spectra were further flux-calibrated and continuum normalized, when applicable \citep[details are provided in][]{lofthouse20}.


The MAGG surveys led to the identification of a large variety of metals (low- and moderate-ions) associated with LLSs \citep{lofthouse23} or selected to be \ion{C}{IV} and \ion{Si}{IV} doublets at $3.0 \le z \le 4.2 $ (Galbiati et. al. submitted), and \ion{Mg}{II} absorbers at $0.9 \le z \le 1.4$ \citep{dutta20}. 

\subsubsection{The QSAGE survey}

The Quasar Sightline and Galaxy Evolution survey \citep{bielby19} is an HST Wide-Field Camera 3 (WFC3) survey 
of 12 quasar fields at redshift $1.2 \le z \le 2.4$ imaged in the near infra-red (NIR; 90\% complete down to $\mathrm{F140W \approx 26\, mag}$) and with
 HST Space Telescope Imaging Spectrograph
 \citep[STIS;][]{kimble98} high-resolution ($\approx 30000$) archival UV spectra.
As for the MAGG survey, the QSAGE quasar fields were supplemented by additional spectroscopy.
Additional medium-to-high resoloution ($\approx12000-24000$) Far-Ultraviolet (FUV) and  Near-Ultraviolet (NUV) data were taken with the HST Cosmic Origins Spectrograph \citep[COS;][]{osterman11, green12} as part of the COS Absorption Survey of Baryon Harbors \citep[CASBaH; e.g.,][]{tripp11}. Together with the STIS data, COS data were reduced using the instrument-specific pipelines, which carried out overscan and bias subtraction, cosmic rays rejection, dark subtraction, flat fielding, spectroscopic wavelength, and flux calibration. 
Finally, supplementary optical high-resolution data ($\approx 40000$) were a compilation of HIRES and UVES spectra retrieved from the Keck Observatory Database of Ionized Absorption toward Quasars \citep[KODIAQ;][]{omeara15,omeara17} and from the Spectral Quasar Absorption Database \citep[SQUAD;][]{murphy19}, respectively \citep[we refer the reader to] [for further details on data reduction]{dutta21}.
The QSAGE survey provides us with \ion{Mg}{II} systems across $z~\approx~0.1 - 1.3$ and \ion{C}{IV} systems across $z~\approx~0.1 - 2.4$ as identified in $S/N~\ge 10$ spectra by \citet{dutta21}.

\subsubsection{The MUDF survey}
The MUSE Ultra Deep Field (MUDF) survey is a MUSE large program (ID 1100.A-0528; PI Fumagalli) targeting a region on the sky containing two bright quasars at $z\approx 3.2$ \citep{lusso19}. 
As a part of the MUDF survey, the MUSE observations were complemented by ancillary UVES high-resolution spectroscopy \citep{dodorico02, fossati19}. As for this work, we make use of the dataset relative to the brighter quasar, which provides us with $S/N \approx 25$ per pixel. Data were reduced with the  UVES  pipeline following a standard reduction process. The reduced spectra were then reformatted with a custom script and input to the ESPRESSO Data Analysis Software \citep[DAS;][]{cupani16} for the final operations of coaddition and continuum fitting. Further details on the data acquisition and data reduction of the MUDF data can be found in \citet{fossati19}. The MUDF provides us with low- and moderate-ion absorbers in the redshift range $z \approx 0.9 - 3.2$.

\subsection{Statistical properties of the absorption line library}
\label{statistics_library}

 MAGG, QSAGE, and MUDF led to a total sample of 688 metal absorption lines. As these data and their fits set the basis for our NMF algorithm with which we aim at tracing and reproducing the lines' intrinsic shapes (see Section \ref {NMF_formalism}), we restricted the sample to only unsaturated metal lines by excluding those profiles for which the continuum normalised flux reaches zero. For each ion, we then selected the strongest transition (highest oscillator strength). However, in case the corresponding profile was saturated, we then selected the weakest transition, subject to the constraint that it was not saturated as well. 
  Less than 1\% of our library consists of medium- low-resolution spectra (e.g. ESI and XSHOOTER data) of single transition lines for which hidden saturation may affect the associated velocity profiles. The remaining sample of medium- low-resolution absorbers consists of doublets and multiple transitions of the same ion for which the effect of hidden saturation is mitigated by MC-ALF, which fits together transitions with different oscillator strengths belonging to the same ion (see Sect.\ref{MCALF_conf}). Finally, since we decouple the column density from the line profile, hidden saturation should not affect building the profile generation library - provided that the shape of the line is not distorted in the core as in the case of evident saturation. 
This resulted in 650 profiles of which we show a small sample and the relative ion contribution in Figure \ref{Sample}.

The moderate-ions (447 profiles) are dominated by \ion{C}{IV} and \ion{Si}{IV} absorbers, while the majority of the low-ions (203 profiles) is represented by \ion{Mg}{II} absorbers.
We also compare the distributions of redshifts, column densities, and $\Delta V_{90}$ values, i.e. the velocity range within which the velocity distribution encompasses 90\% of the optical depth of the line, relative to the low- and moderate-ions classes (cyan and red sample in Figure \ref{Sample_dist}). 
On average the low-ion distribution peaks at lower redshifts, ${z_{\mathrm{l}}} = 2.4 \pm 1.1$, than the distribution traced by the moderate-ions, ${z_{\mathrm{m}}} = 3.4 \pm 0.4$. On the other hand, the column density distributions peak at similar values although they are characterised by a different dispersion, i.e.  $\mathrm{\log(N/cm^{-2})_{l} = 13.1 \pm 0.9}$ and $\mathrm{\log(N/cm^{-2})_{m} = 13.4 \pm 0.7}$.
Finally, when the $\mathrm{\Delta V_{90}}$ distribution is plotted separately for the low- and intermediate-ions, the highly ionised species can show broader line widths.


\section{MC-ALF: the Monte-Carlo Absorption Line Fitter code}\label{Voigt_fitting}

To extract the wealth of information on the kinematic, chemical, and ionization conditions of 
the gas probed by absorption line systems, it becomes necessary to model the spectral features. 
Albeit non-parametric techniques exist \citep[e.g. apparent optical depth, AOD; ][]{savage91}, Voigt fitting has become the main modeling technique to extract the lines' physical properties.
 Decomposing absorption line profiles into Voigt components can be an expensive task, particularly because of the degree of subjectivity in setting initial conditions. Alternative approaches to Voigt profile fitting that used $\chi^{2}-$based codes \citep[e.g.][]{fontana95,dave97,cooke14,carswell14,krogager18} have been sought, e.g. by using Bayesian techniques. These techniques have the advantage of being relatively less computationally expensive in cases where multiple absorption component fitting is needed. Moreover, by sampling the posterior distribution of the parameters' values their uncertainties and degeneracies can be better constrained.

Within this framework, one example is \texttt{BayesVP} \citep{liang17} which models Voigt profiles and generates posterior distributions for the column density, Doppler parameter, and redshifts of the corresponding absorber. However, it is based on an affine-invariant MCMC sampler that does not easily converge in a high dimensional parameter space, thus resulting in computationally expensive runs when the initial conditions or the number of free parameters are not known. To obviate this problem, one can resort to nested sampling \citep{skilling06} which provides complete statistical information and makes it possible to efficiently carry out model comparison via the Bayesian evidence.

 In this work, we present the technical details and quality assurance tests of a new Bayesian fitting code first introduced in \citet{fossati19} and dubbed the Monte-Carlo Absorption Line Fitter (MC$-$ALF).
MC$-$ALF has four innovative features compared to other absorption line fitting codes: i) it requires minimal input from the user as no initial conditions are given but only the allowed range of the parameters is required; ii) using Bayesian statistics the final result provides the full posterior distribution for each parameter and their covariance matrix, leading to an optimal statistical description of the data; iii) it samples the multidimensional likelihood space using {\sc PolyChord} \citep{Handley15}, a nested sampling algorithm that has the best performance for high-dimensional parameter spaces with multiple degeneracies between parameters, as it is the case of the multi-component Voigt parametrization of complex absorption profiles; iv) it is naturally adaptive, shown to accurately retrieve all the information of both high- and low-resolution profiles with execution times that scale with the complexity of the profile (typically related to the instrument resolution and data $S/N$. See Sections \ref{MCALF_conf} and \ref{MC$-$ALF_test} for more details). All these characteristics make MC$-$ALF ideally suited to study in an automatic fashion big-data samples from large spectroscopic surveys where the combination of moderate resolution and $S/N$ reduces the need for complex (and expensive to compute) absorption models.

\subsection{Formalism for Voigt profile fitting}

The analytic model at the basis of MC$-$ALF is the canonical combination of multiple Voigt functions that are used to describe line profiles of any complexity. 
The absorption line arising from a transition $i$ of an ion can be described by the optical depth of the transition, $\tau_{i} (\nu)$, which is determined by the column density of the ion, $N$, along with a set of atomic parameters describing the line strength, $f_i$, the damping constant, $\Gamma_i$, and the resonance frequency, $\nu_i$, i.e. 
\begin{equation}
\tau_{i} (\nu) = {N} s_{i} \phi_{i}(\nu),
\label{optical_depth}
\end{equation}
 where $s_{i}$ is the frequency-integrated absorption cross-section given by
\begin{equation}
s_{i} = \frac{\pi e^{2}}{m_{e} c}f_{i},
\end{equation}
with $e$ the electron charge, $m_e$ the electron mass, and $c$ the speed of light. Finally, the frequency-dependent line profile for a single component is
\begin{equation}
\phi_{i}(\nu) = \frac{H(u_{i},a_{i})}{\Delta \nu_{i} \sqrt{\pi}},
\end{equation}
with the Voigt function
\begin{equation}
\label{voigt_f}
H(u_{i},a_{i}) = \frac{a_{i}}{\pi} \int_{-\infty}^{+\infty} dy \frac{{\rm exp}(-y^{2})}{(u_{i}-y)^{2} + a_
{i}^{2}},
\end{equation}
where $a_{i}= \Gamma_{i} / 4\pi \Delta\nu_{i}$, $u_{i}=(\nu - \nu_{i})/\Delta \nu_{i}$ is the re-scaled frequency, $y = v/b$ is the velocity in units of the Doppler parameter, $b = (2kT /m + \xi^{2})^{1/2}$, which is given by the gas temperature, $T$, the element mass, $m$, and the turbulent velocity, $\xi$. Finally, $\Delta \nu_{i} = \nu_
{i,0} b/c$, with $\nu_{i,0}$ the transition rest-frame frequency.
 Thus, the Voigt function is the convolution of the Gaussian line broadening due to thermal and turbulent motion with the Lorentzian contribution from natural line broadening and it can be separated from the normalization factors, $N$ and $s_{i}$, intervening in Equation \ref{optical_depth} to model the optical depth of the considered transition.

By summing over all the $\mathrm{N_V}$ Voigt components, the resulting transmitted flux, $I_{i} (\nu)$, of a background source with intensity, $I_{0}$, is given as
\begin{equation}
\label{model}
    I_{i}(\nu) = I_{0} \rm{e}^{-\sum_{j}^{\rm{N_V}}\tau_{ij}(\nu)}.
\end{equation}

Equation \ref{model} defines our model and its free parameters are the column density, the Doppler parameter, the redshift, and the number of components, i.e. $\theta = \{{\rm N},b,z,\rm{N_V}\}$. 


\subsection{Bayesian inference and posterior sampling }
 Within the Bayesian inference framework the posterior distribution, $\mathcal{P}$, is proportional to the product between the likelihood, $\mathcal{L}$ and the prior probability distribution functions (PDFs), $\pi$, so that

\begin{equation}
\mathcal{P(\theta)} \propto \mathcal{L}(\theta) \times \pi( \theta)
\label{posterior}
\end{equation}

 is proportional to the product between the probability of observing the data given a specific set of parameters and their prior distributions. Specialized to our case, the data is represented by the measured flux, $F$, in a spectral interval. Then we can write:

\begin{equation}
\mathcal{L}(\theta) = \prod_{i} l_{i}(\theta|F_{i}),
\end{equation}

were $l_{i}$ is the likelihood relative to an individual pixel $i$, i.e.

\begin{equation}
    l_{i} = \frac{1}{\sqrt{2\pi \sigma_{i}^{2}}} \rm{exp}\left[{\frac{\bar{F}_i(\theta) - F_{i}^2}{2\sigma_{i}^2}}\right],
\end{equation}

with $\sigma_{i}$ the flux error.

The likelihood space is sampled via the nested sampling algorithm {\sc PolyChord}.  The sampling starts with a large number of live points ($n_{\rm live}$) within a region of the parameter space sampled by the prior distribution. These points are sequentially updated so that those with the smallest value of the posterior density are eliminated (termed dead points) and are replaced by a new live point, again drawn from the prior, whose likelihood is larger than that of the point that was discarded. To generate new points {\sc PolyChord} uses the so-called slice sampling where new live points are generated by taking a random slice through the parameter space that includes the current live point, and randomly generating new points until one with a higher likelihood is found. The process is then repeated with the new point and a slice in a new random direction, for a user-defined number of repetitions ($n_{\rm repeat}$). The length of this chain of repetitions should be large enough so that the final live point is decorrelated from the start point.


\subsection{Model comparison }
\label{model_comparison}
 Metal absorbers can be characterised by complex profiles where line blending can make it difficult to retrieve the number of Voigt components that better define the observed profile. This, together with the fact that we aim at sampling a large space of parameters, yields the necessity of choosing between competing models. In turn, this capability obviates the need for the user to specify a set of initial conditions or strong priors for the parameters. 
In a pre-release version of MC$-$ALF code used by \citet{fossati19}, the number of Voigt components was kept as a fixed parameter at each fit iteration and multiple fits with an increasing number of components were performed to decide on the best decomposition model using the Akaike Information Criterion \citep[AIC,][]{Akaike74}. To improve the code performance, a non-negligible aspect for deploying this code in large surveys, we have refactored the code to include the number of components in the likelihood calculation, so that a single fit can be performed keeping the number of components as a free parameter. Thus the algorithm is terminated once the improvement in the likelihood is some small fraction of the currently calculated once. Moreover, this version has the added value of providing posterior distributions of the number of components which can be useful in the case of highly complex profiles.

\subsection{MC-ALF configuration file}
\label{MCALF_conf}
A MC$-$ALF configuration file has three main blocks:{\bf\texttt{ input}}, {\bf \texttt{components}}, and {\bf \texttt{pcsettings}}, with which the user defines the input information, the parameters for the components to be fitted, and the setting of the {\sc polychord} solver through their attributes.
In {\bf \texttt{input}} the main information the code requires is the spectral data to fit. This is an ascii table with three columns providing the wavelength in \AA, the continuum normalised flux, and its error. There is no preferential order with which the columns must be organised as long as this information is provided in the \texttt{coldef} attribute. The user will then have to specify the transitions to fit (only atomic transitions belonging to the same ion can be fit together), following a naming convention that sees the ion name followed by its rest-wavelength in \AA\, and separated by a white space. 

\begin{figure}
    \centering
    \includegraphics[width=\columnwidth]{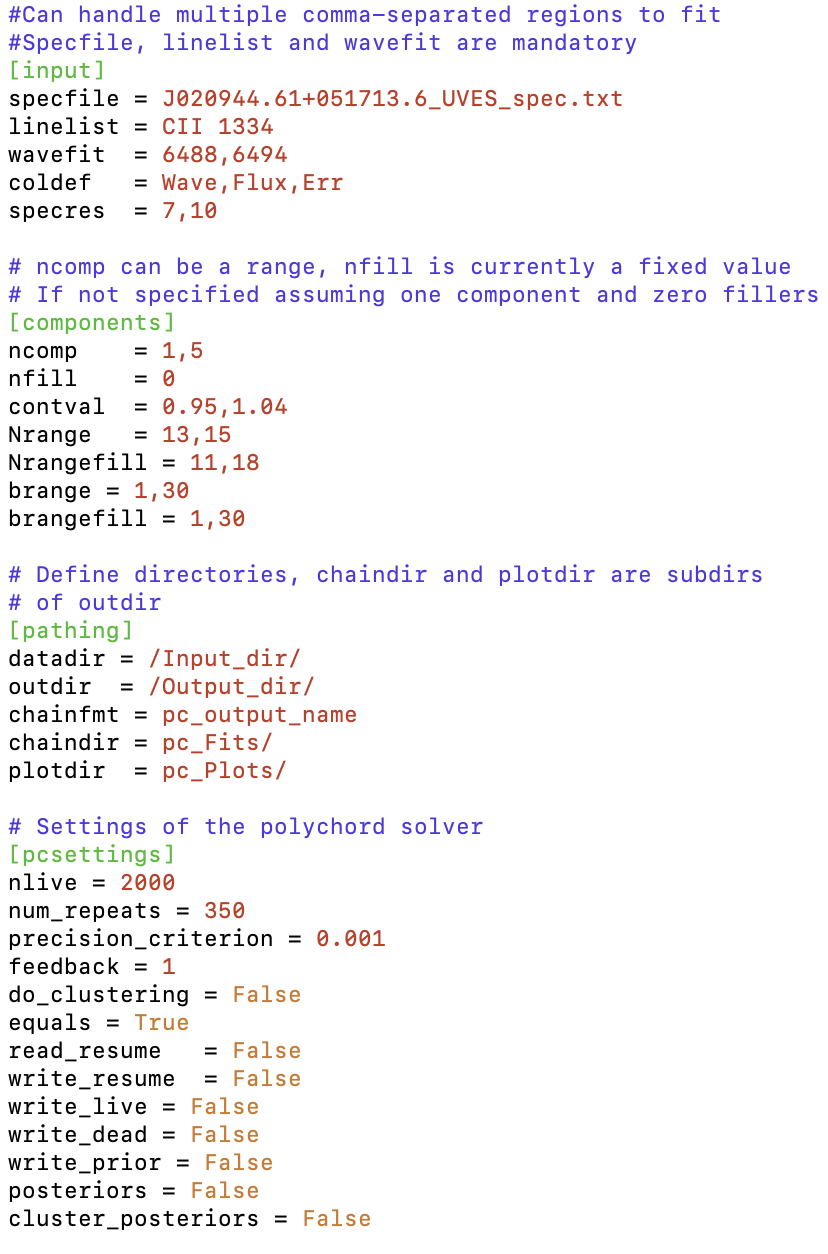}
    \caption{Example of an MC$-$ALF configuration file. In this example, the ion to fit is a \ion{C}{II} ion in a high-resolution spectrum as specified in the \texttt{linelist} and \texttt{specres} attributes in the {\bf\texttt{input}} block. The range of the fitted parameters, i.e. the column density ($N$, in log units), the Doppler parameter ($b$ in km~s$^{-1}$), and the redshift of the transition are specified in the {\bf\texttt{components}} block. If they are not provided, the code assumes default values. The {\bf \texttt{pcsetting}} block controls the parameters of the {\sc PolyChord} algorithm (see text for more details)}.
    \label{conf_file}
\end{figure}

Next, the user will provide the wavelength range (or disjoint ranges) to fit the data with Voigt components. These are described by their column density ($N$ in log units of $\rm cm^{-2}$), the Doppler parameter ($b$ in $\mathrm{km\, s^{-1}}$) and the redshift of the transition. The number of Voigt components is an additional free parameter in the fit and the user will specify the range to be explored via the \texttt{ncomp} attribute. Similarly, the range of $b$-parameter values to be fitted can be passed as \texttt{brange}. If required, it is possible to include a user-defined number of ``filler'' Voigt profiles designed to describe absorption lines arising from blends of different ions at different redshifts in the wavelength range being fit and controlled by the \texttt{nfill} attribute. The range of column density, and $b$-parameter values for the ``fillers'' is then passed via  \texttt{Nrangefill} and \texttt{brangefill}, respectively. Note that, while the dynamic range of each of the free parameters can be specified in the code configuration file, reasonable default values are provided to the code. \\
Finally, the user can control the parameters of the {\sc PolyChord} algorithm directly in the {\bf \texttt{pcsettings}} block, defining the number of live points (\texttt{nlive}) and the number of slices (\texttt{num\_repeats}) at each iteration, therefore, balancing execution time and the likelihood accuracy. A more detailed description of these parameters can be found in \citet{Handley15}. An example of an MC$-$ALF configuration file is shown in Figure \ref{conf_file}.

We note that the code's upgrade of including the number of Voigt components as free parameters (see Sect. \ref{model_comparison}) has improved the execution time by a factor of $5-10$ so that, in its default configuration (\texttt{ncomp} = 1-15, \texttt{nlive} = 500, \texttt{num\_repeats} = 50) MC$-$ALF takes $\sim1.3$ total CPU hours to run on recent Intel processor and to model multiple-components, high-resolution spectra. As the models are expected to be simpler for lower-resolution and lower-$S/N$ data, the code can be optimized for speed by reducing the interval of components to be considered and by reducing the \texttt{nlive} and 
\texttt{num\_repeats} (for typical WEAVE-like data we set these to (\texttt{ncomp} = 1-3, \texttt{nlive} = 350, and \texttt{num\_repeats} = 50, having the profiles fully analysed in tens of seconds on a single core machine).

\subsection{MC-ALF output}
\begin{figure}
	\centering
	\includegraphics[width=\columnwidth
	]{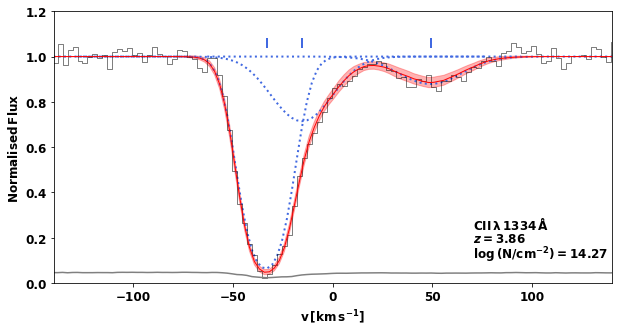
	}\\
	\includegraphics[width=\columnwidth
	]{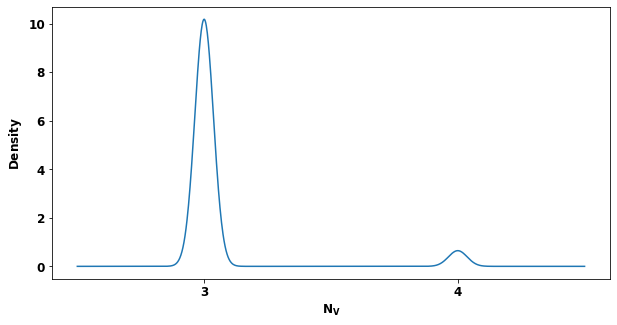
	}
\caption{\textit{Top}: The results of the fit procedure on a \ion{C}{II} absorber. The solid red line is the median model of all the posterior samples, while the red shaded region represents the $\pm 1\sigma$ uncertainty on the model profile. The blue dotted lines represent individual Voigt components, centered at the velocities highlighted by the blue ticks. \textit{Bottom}: The kernel density distribution of the number of Voigt components in the posterior samples. Nearly all the samples fit the data with three Voigt components. Note that the number of components is an integer in our fitting model.}
    \label{MC$-$ALF_fit}
\end{figure}

\begin{figure}
	\centering
	\includegraphics[width=\columnwidth
	]{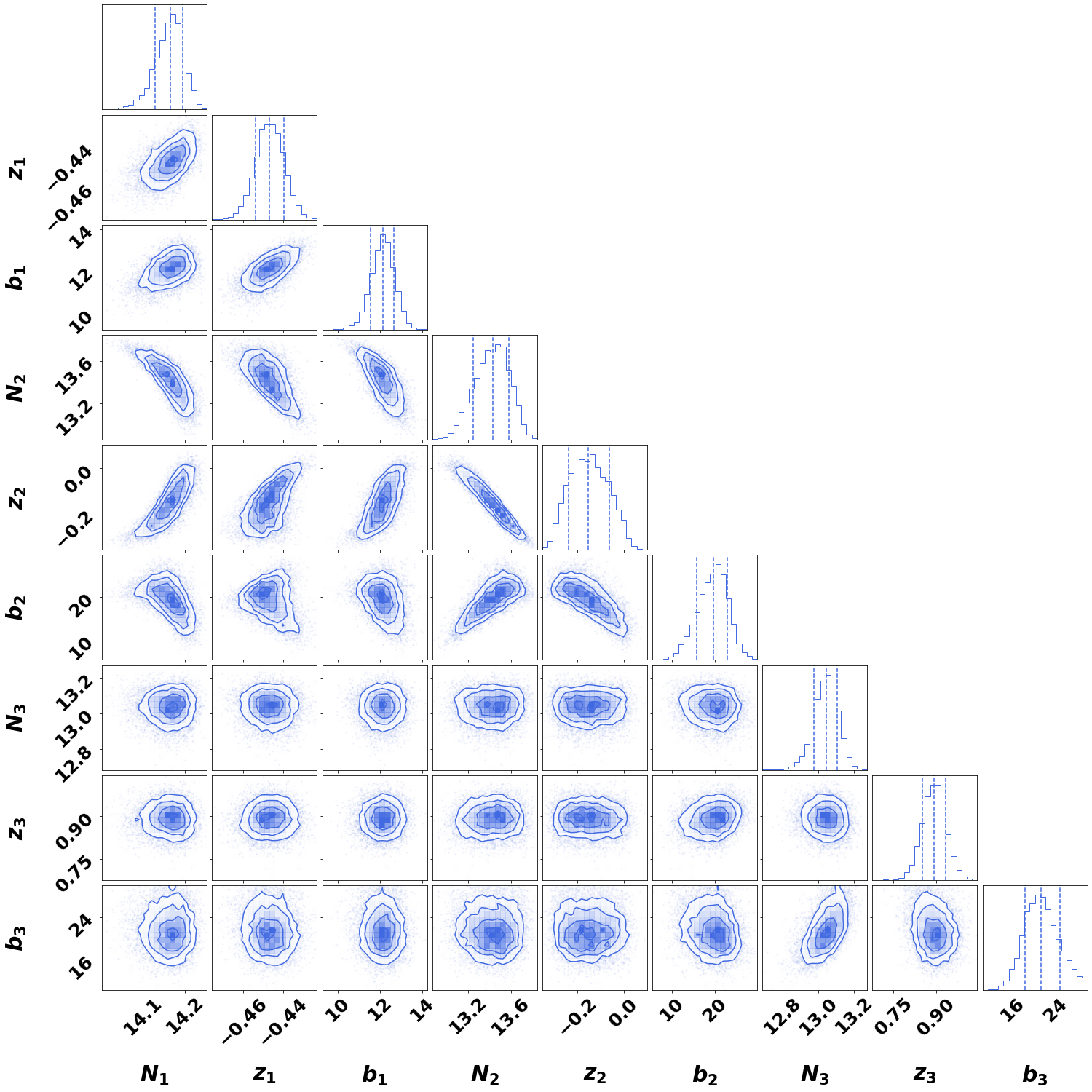}
\caption{Corner plot of the posterior samples for the MC$-$ALF fits with three Voigt components of a \ion{C}{II} absorber at $z\approx 3.86$. The contour panels show the posterior distribution of pairs of free parameters, while the histograms show the 1D posteriors of individual parameters. The dashed vertical lines correspond to the 16$^{\rm th}$, 50$^{\rm th}$, and 84$^{\rm th}$ percentiles of the distributions, respectively. Units for $b$-parameters are $\mathrm{km\, s^{-1}}$, column densities are in $\log{\mathrm{(N/cm^{-2})}}$. }
    \label{MC$-$ALF_cornerplot}
\end{figure}

 When the fit has converged, the MC$-$ALF output is saved in an ascii file that will list the following columns for each posterior sample: the components' weight, the total evidence, the best likelihood, the fitted continuum, and spectral resolution (if they are free parameters), and the values of the fitted parameters, i.e, the column density, the redshift, and the $b$-parameter. As an example, Figures \ref{MC$-$ALF_fit} and \ref{MC$-$ALF_cornerplot}  show the model fits obtained with MC$-$ALF. The fit is run on a \ion{C}{II}$\lambda 1334$ \AA\, absorber (high-resolution spectrum, $8\, \mathrm{km\, s^{-1}}$) with a value for the column density $\rm{\log{(N/cm^{-2})}} = 14.3$.
We show that the best model reproduces the data mainly with three Voigt components, although there is a non-zero probability that a four-component model can also be compatible with the data (Figure \ref{MC$-$ALF_fit}). Moreover, the full posterior distribution is saved and can be used for further processing and analysis by the users. For example, the corner plot in Figure \ref{MC$-$ALF_cornerplot} shows the covariance and distribution of Voigt parameters for each component ($N$, $z$, and $b$).

\subsection{Code validation and quality assurance tests}
\label{MC$-$ALF_test}

\begin{figure*}
	\includegraphics[width=7.5cm]{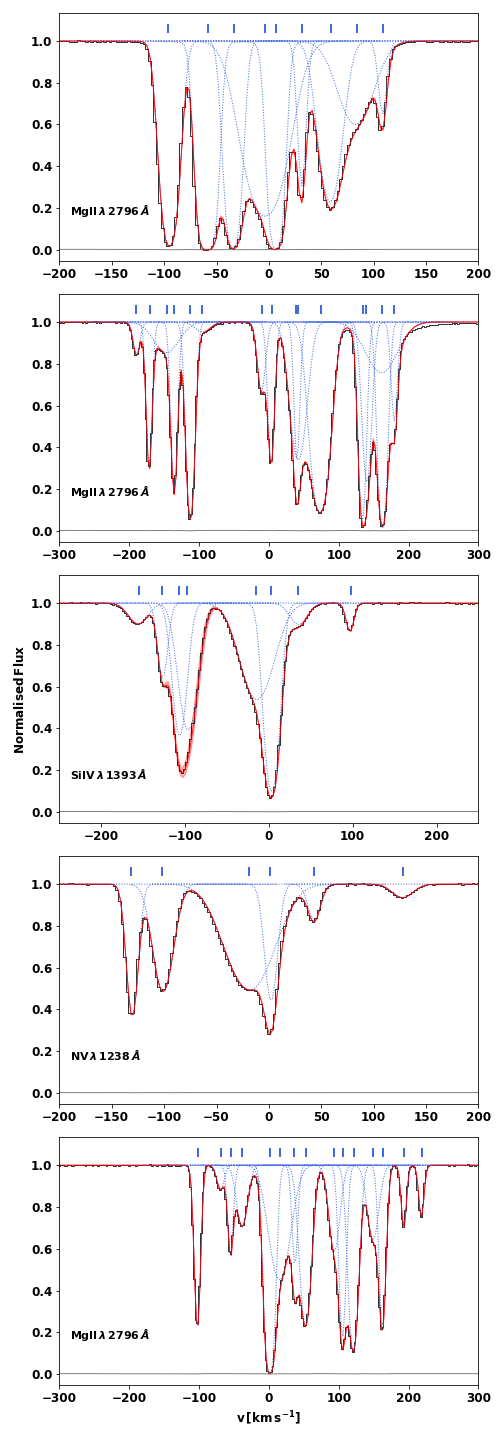}
	\includegraphics[width=7.5cm]{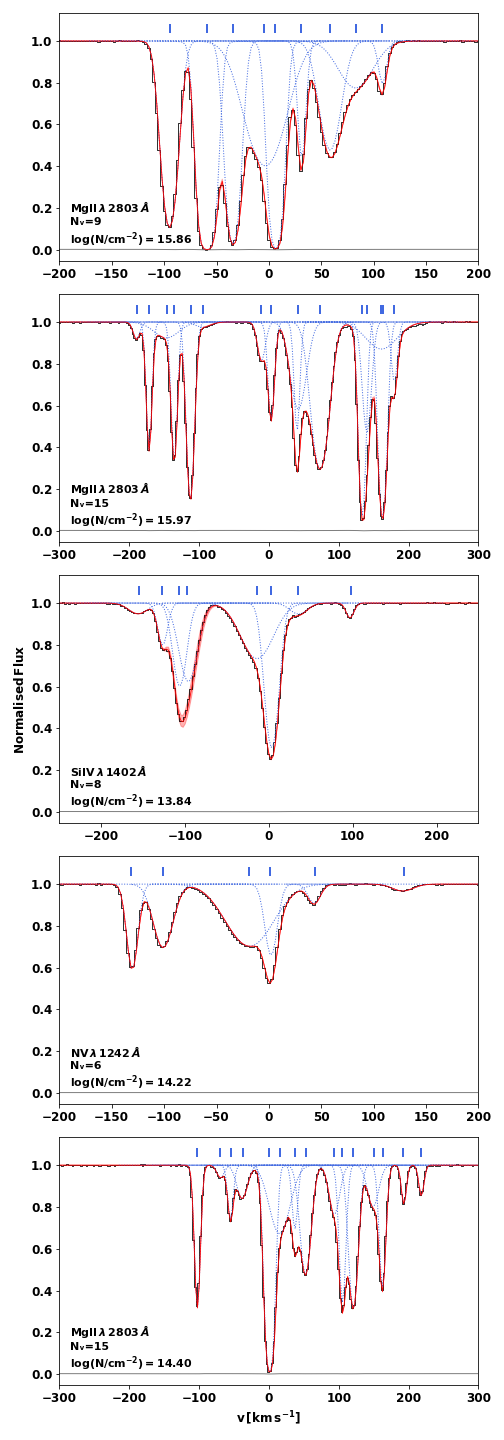}
    \caption{Simulated profiles (black) used to test the Voigt Component fitting routine. The profiles are characterised by different column densities and number of Voigt components as given in the legend. For simplicity, we only show the case for $S/N =500$, with the $1\sigma$ sigma array in gray. For each profile, the MC$-$ALF fit is shown as a solid red line with $\pm 1\sigma$ uncertainty as a shaded area. The dotted blue lines represent individual Voigt components centered at the velocities highlighted by the blue ticks. }
    \label{test_profiles}
\end{figure*}

\begin{figure*}
	\centering
	\includegraphics[trim={2.5cm 2.5cm 5cm 4.5cm}, clip,width=16.5cm]{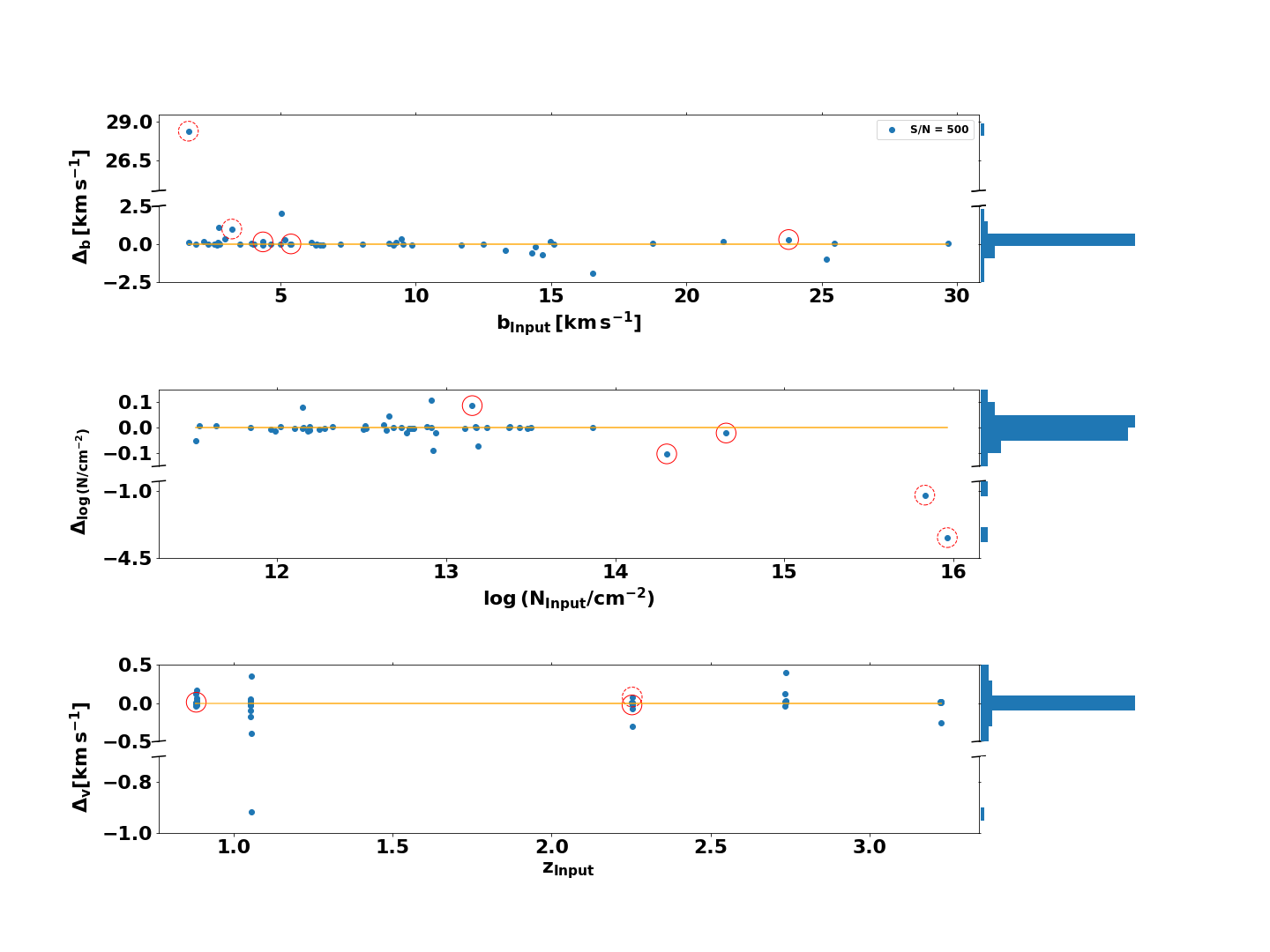}
    \caption{$b$-parameters, column densities, and velocity accuracy for the retrieved sample of Voigt components when the profiles with $S/N=500$ are analysed. Continuous and dashed red circles identify saturated lines and lines with $\mathrm{\log{(N/cm^{-2})} > 15}$, respectively.  The largest deviating point in the retrieved $b-$parameter corresponds to one of these lines.The histograms of the residuals are plotted on the right.}
    \label{vc_testing1}
\end{figure*}

\begin{figure*}
	\centering
	\includegraphics[trim={3.5cm 3.5cm 3cm 3cm}, clip, width=16.5cm]{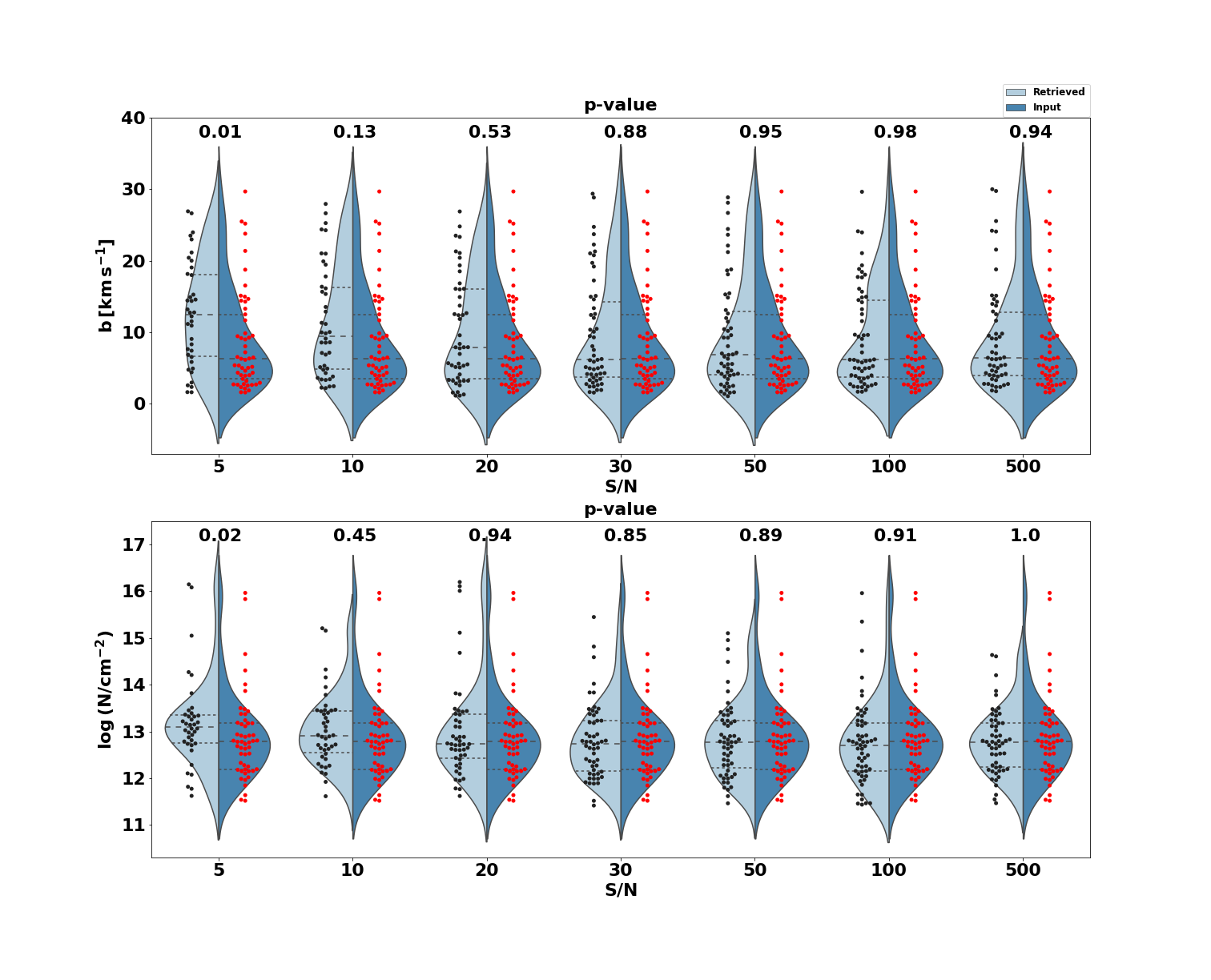}
    \caption{ Violin plots comparing the probability density of $b$-parameters and column densities as traced by the input (light blue) and retrieved (dark blue) samples as a function of the $S/N$ of the input spectra. In each violin, the horizontal central dashed line is the median and the dotted lines are the 25\% and 75\% quartiles. The distributions are determined from the entire sample of Voigt components as depicted by their relative swarm plots (dots). MC$-$ALF recovers the total input distributions ($p-$value $> 0.1$) for $S/N \ge 10$. At lower $S/N$ the code must be tested relative to the integrated values of the fitted parameters. }
    \label{vc_testing2}
\end{figure*}

We test MC$-$ALF capabilities by analysing a set of 5 synthetic profiles mimicking \ion{Mg}{II}, \ion{Si}{IV}, and \ion{N}{V} systems, characterised by different total column densities, $13.8 \le \mathrm{\log{(N/cm^{-2})}}\le16.0$, and number of Voigt components, $6 \le N_{V
} \le 15$, at different redshifts, and for which we have {\it a-priori} knowledge of the Doppler parameter and column density values relative to each Voigt component (Figure \ref{test_profiles}). The profiles have been created to reproduce the performance of UVES/VLT spectra, i.e. they are characterised by a resolution of $8\, \mathrm{km\, s^{-1}}$ with a pixel sampling of $2.5\, \mathrm{km~s^{-1}}$. Finally, a Poisson noise component is added in each profile, $\mathrm{\sigma = \sqrt{\sigma_{source}^2 + \sigma_{sky}^2}}$, with $\mathrm{\sigma_{source}}$ so that 
the MC$-$ALF performances could be tested as a function of different $S/N$ ratio (per pixel) with respect to the continuum of the background source, namely $S/N = 5$, 10, 20, 30, 50, 100 and 500, and the sky noise, $\sigma_{sky}$, so that the continuum dominates by a factor 4 on the sky signal already at $S/N =10$.  

Next, we model these profiles with MC$-$ALF and the results are shown in Figure \ref{vc_testing2} where we compare the probability densities of the input and retrieved distributions of $b$-parameters and column densities as a function of the different $S/N$ of the input spectra. The Kolmogorov-Smirnov (KS) test $p$-value scores (given on top of the respective distributions) show that Doppler parameter estimates are more sensitive to the quality of the analysed spectra compared to that of the column density estimates. Nonetheless, MC$-$ALF is able to recover the total input distributions ($p-$value $> 0.1$) of both the $b$-parameters and column densities already at $S/N=10$. At lower $S/N$ ratios, some discrepancies are found when analysing individual components because it becomes increasingly difficult to accurately match, in a statistical sense, single input-$vs$-retrieved Voigt components (leading to $p-$value scores $< 0.1$).
Hence, these discrepancies reflect a mismatch in components rather than inaccuracies in  MC$-$ALF to recover values. We find that the retrieved fraction of the $\mathrm{N_V}$ components is on average $<f^{\mathrm{{N_V}}}> =$ 0.77, 0.83, 0.88, 0.88 for $S/N$ = 5, 10, 20, and 30, respectively, while at higher $S/N$ MC-ALF retrieves all the input components (except for single high-density components that MC-ALF may split in two; see below). Thus, in the low $S/N$ regime, it is better to test the code capabilities relative to the total (integrated) values of the fitted parameters. When these are considered MC$-$ALF successfully retrieves the input total distribution with a mean relative error of $<\delta^{\mathrm{\log{N}}}> = 0.12,0.09,0.11,0.08$ at $S/N =5,10,20$, and 30. In Section \ref{conclusions} we provide additional tests for thousands of simulated profiles in the low $S/N$, low-resolution regime.

Focusing on the $S/N=500$ test, we compare 
the recovered values of $b$-parameters, column densities (in $\log{\mathrm{(N/cm^{-2})}}$), and redshifts (the latter being converted to $\Delta \mathrm{V} = c\frac{z_{\mathrm{Ret}} - z_{\mathrm{Input}}}{1+z_{\mathrm{Input}}}$, with $c$ the speed of light ) against their input values (Figure \ref{vc_testing1}). 
For the three distributions, we find mean relative errors of $<\delta> = 0.4, 0.007,\, \mathrm{and}\, 0.62\times 10^{-6}$.  We note that the errors for the $b-$parameters and column densities are dominated by the errors associated with single Voigt components with $\mathrm{\log{(N/cm^{-2})}} > 15.0$. When these are excluded from our computation the relative errors for the two distributions drop to $<\delta> = 0.03,  0.002$, respectively.  These high column density components are also responsible for MC$-$ALF to find in output one additional Voigt profile for the \ion{Mg}{II} absorber with $\mathrm{\log{(N/cm^{-2})}} = 15.97$ (second-row panels in Figure \ref{test_profiles}). This phenomenon is the result of the decomposition of the single Voigt profile in a narrow component that best describes the high optical depth regime and a broad component that best fits the wings. In Figure \ref{vc_testing1} this spurious detection is responsible for the most discrepant $\Delta b$ value.

\begin{figure*}
    \centering
    \includegraphics[width=\textwidth]{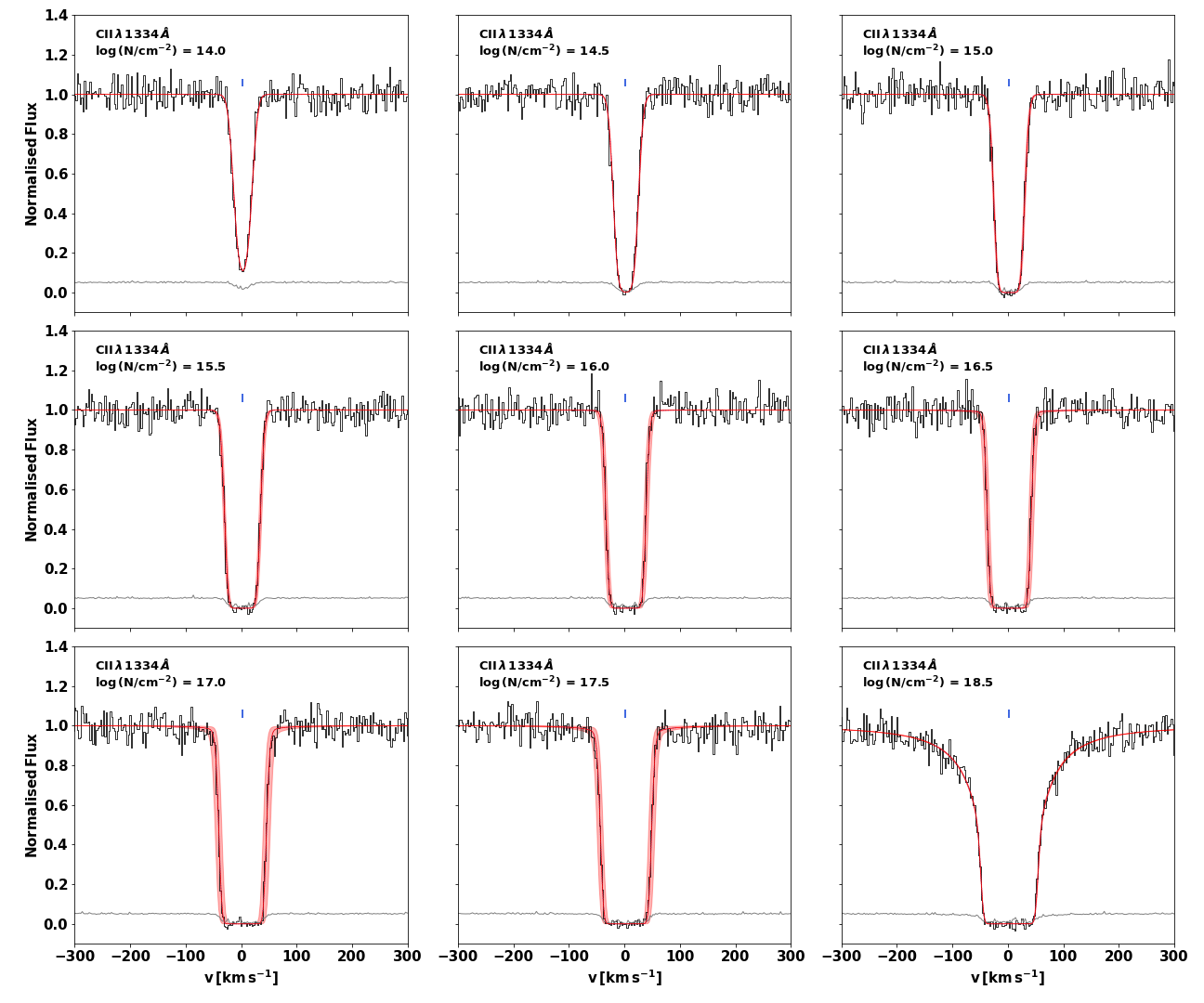}\\
    \caption{Simulated \ion{C}{II} profiles (normilised flux in black and error in gray)  with $b = 15\, \mathrm{km\, s^{-1}}$ and $S/N=20$. From top to right the profile is shown with higher column density values as given in the legend. For $\mathrm{log{(N/ cm^{-2})}} \ge 14.5$ the profiles are saturated and for $\mathrm{log{(N/ cm^{-2})}} = 18.5$ the damping wings of the Lorentzian become significant compared to those of the Gaussian contribution. For each profile, the MC$-$ALF fit is shown as a solid red line with $\pm 1\sigma$ uncertainty as a shaded area and centered at the velocities highlighted by the blue ticks.}
    \label{test_sat_profiles}
\end{figure*}

 Finally, in Figure \ref{vc_testing1} we mark as red-empty dots the saturated Voigt components, i.e. with flux density levels reaching zero in the normalized spectra, for which the column density value may no longer be estimated with a few percent accuracy. We then additionally tested the performances of MC$-$ALF in presence of saturation. As before, the test is carried out on a UVES/VLT-like synthetic profile. The absorber is a single Voigt component of \ion{C}{II} at $\lambda 1334$ \AA\, with a $b-$parameter fixed at $b = 15\, \mathrm{km\, s^{-1}}$ and column density values  $14 \le \mathrm{log{(N/ cm^{-2})}} \le 18.5$. The profiles, shown in Figure \ref{test_sat_profiles} in the case of $S/N=20$, are saturated for $\mathrm{log{(N/ cm^{-2})}} \ge 14.5$ and for $\mathrm{log{(N/ cm^{-2})}} = 18.5$ (an extreme value useful for testing) the damping wings of the Lorentzian become significant compared to those of the Gaussian contribution (see Equation \ref{voigt_f}).

\begin{figure}
    \centering
    \includegraphics[width=\columnwidth]{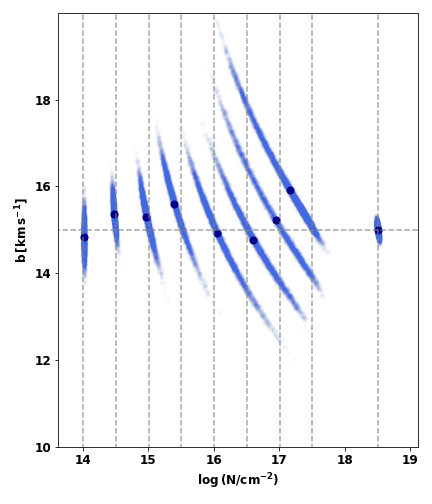}
    \caption{$b\, vs\,  \mathrm{log{(N/ cm^{-2})}}$ plot to test MC$-$ALF fits on a saturated \ion{C}{II} absorber. Blue dots trace the full posterior samples provided in output by MC$-$ALF. The dotted gray lines show the simulated $b$ and column density values. In presence of saturation, however, when the damping wings are not yet significant, the column density estimate is on average a lower limit with a relative error of $<\delta^{\mathrm{\log{N}}}> = 0.006$.  When the damping wings contribution becomes significant, in our example for $\mathrm{log{(N/ cm^{-2})}} > 18$, the fit returns accurate estimates of the column density.}
    \label{test_sat}
\end{figure}

The impact of saturation on the fits is shown in Figure \ref{test_sat} as $b\, vs\,  \mathrm{log{(N/ cm^{-2})}}$ plot. In this figure, the blue dots trace the full posterior samples provided in output by MC$-$ALF and the dotted gray lines show the simulated $b$ and column density values. When the line is not saturated, in our example for $\mathrm{log{(N/ cm^{-2})}} = 14.0$, the retrieved columns density and $b-$parameter is a sensitive measure of their true values. Moving towards the saturated regime, the column density estimate is a lower limit with an average relative error of $<\delta^{\mathrm{\log{N}}}> = 0.006$. The $b-$parameter is yet well constrained with an average relative error of $<\delta^{b}> = 0.03$. Finally, for $\mathrm{log{(N/ cm^{-2})}} = 18.6$ the optical depth in the damping wings becomes significant and the fit returns accurate estimates of the column density, as expected.

\section{NMF-PM: the NMF Profile Maker code}
\label{NMF_formalism}
 We now introduce the second tool we present in this paper, NMF Profile-Maker (NMF$-$PM). 
In what follows we first outline the data standardization steps we followed to prepare our library of metal profiles for the NMF analysis. Afterwards, we present a brief overview of the NMF formalism and outline the details of how we built a statistically robust process of NMF reconstruction and simulation. Finally, we show how the results of this analysis are used to build the NMF$-$PM python module, a metal absorber profile maker which we make publicly available.

\subsection{Data standardization}\label{data_standardization}

 NMF is an alternative approach to dimensionality reduction (e.g., to principal component analysis or PCA) where it is assumed that the data can be decomposed (or transformed) into non-negative components.  Despite its desirable properties ( it automatically extracts sparse and meaningful features from a set of non-negative data vectors), the NMF fitting requires the data to be normalised and regularised for an unbiased decomposition. 
By using MC$-$ALF on our library of absorbers, we obtain infinite $S/N$ Voigt models of absorption profiles of different strengths and with a range of velocity distributions (see Figure \ref{Sample}). In particular, our library consists of 650 unsaturated metal profiles of which we aim at reproducing their intrinsic shape (Section \ref{statistics_library}).
To prepare and standardise the data for the NMF decomposition, we adopt the following two-step procedure.  

As a first step, we need to determine the rest-frame velocity of the profiles and shift them to a common velocity frame. 
For this, we can use the model Voigt components to re-sample the profiles at a resolution of $1\, \mathrm{km\, s^{-1}}$ and to transform them to a common rest-frame velocity centered at $0\, \mathrm{km\, s^{-1}}$.  As the profiles may exhibit several Voigt components of different strength, here we define the zero velocity as $\Delta\, V_{50}$, i.e. the value at which the velocity distribution encompasses 50\% of the optical depth of the line.

The second step involves the recovery of the optical depth and the normalisation of the line profiles. Rather than considering the transmitted flux, we elect to describe profiles in terms of their optical depth, $\tau(\nu)$, for which non-negativity is inherent to the data being considered and the normalisation step is more straightforward. As recalled in Equation~\ref{optical_depth},  $\tau(\nu)$ is the product of the ion column density, $N$, the frequency integrated absorption cross-section, $s$, and of the velocity profile. 
Thus, the normalization of the optical depth by the product $N\times s$ allows us to retrieve the line intrinsic profiles, $\phi(\nu)$, without carrying the added complexity of individual column densities and oscillator strengths of different absorbing ions, and to focus on the line shape as the only general property we wish to describe and reproduce in the mock-making step.
 Once velocity profiles are generated, the full absorption line systems can then be recovered by multiplying back the desired column density and strength of an ion. The imperfect approximation we are introducing at this step is to separate the correlation between an ion and its Doppler parameter, due to the atomic mass dependence. This approximation fails in the limit of single lines that are thermally broadened, but it holds for the majority of the profiles where turbulence and the combination of multiple components determine the line shape. Finally, as the profiles used to train and test the NMF algorithm are models obtained via runs of MC-ALF on our set of observed data, very small structure is lost at moderate resolutions compared to high-resolution modes, so our modeling performs best for resolutions that are comparable to the lowest one in our library, i.e. X-shooter-like, and caution should be taken when applying this model to particularly high spectral resolutions.
 

\subsection{Application of the NMF method}\label{subsec:application_NMF}

\subsubsection{Overview of the formalism and general concepts}

The NMF formalism assumes that a non-negative dataset of $n$ samples and $v$ features can be approximated by the (dot) product of two non-negative matrices. 
\begin{equation}
   \bf {D} \approx \bf{XC} 
    \label{NMF}
\end{equation}
where ${\bf D} \in \mathbb{R}^{n \times v}$ is the matrix representation of the original data. Matrix $\bf X$ has the shape $n \times m$, where $m$ is the number of reduced features in NMF space. The matrix $\bf C$ has the shape $m \times v$ and represents the coefficient matrix of the $m$ reduced features, or in other words a representation of the new reduced features in the original feature space. Thus, via NMF, we generate a low-dimensional encoding of a high-dimensional space.
From Equation~\ref{NMF}, it follows that each row in the matrix {\bf D} (each sample) is a linear combination of the row vector in the matrix {\bf X} with coefficient vectors supplied by the matrix {\bf C}, i.e.
\begin{equation}
   {\bf d_{i}} = \sum_{j=1}^{m} {\bf x_{j}} {c_{jv}}\:.
    \label{NMF_spectrum}
\end{equation}
Thus, NMF recasts an original vector onto new component axes of \textit{latent features}, $x_{j}$, and the projections onto such an NMF space are given by the vectors in $C$.

Specialized to our application, we have $n$ line profiles characterised by $v$ velocities, which collectively can be represented by a matrix $\bf Q$ of dimension $n\times v$. We wish to reduce the dimensionality of the problem and assume that each of these profiles can be represented by $m$ features where $m < v$. We apply NMF to $\bf Q$ and obtain two matrices $\bf X$ and $\bf C$ whose matrix multiplication is represented by $\bf{R}$:
\begin{equation}
   \bf {Q} \approx \bf{R} = \bf{XC}. 
    \label{NMF_applied}
\end{equation}
We find $\bf R$ such that it is the closest representation of $\bf Q$. The decomposition works by minimizing the squared Frobenius norm (i.e. a generalization of the Euclidean norm to matrix algebra) between ${\bf Q}$ and the matrix product ${\bf X}{\bf C}$. In particular, our NMF fit implements a coordinate descent solver, i.e. an iterative process that successively updates the fitted parameters until convergence is reached. 

Once $\bf{R}$ is obtained we can further create a synthetic set of profiles by randomly assigning NMF latent features from their retrieved distributions in ${\bf X}$ and then carrying out the linear combination as in Equation \ref{NMF_spectrum}, i.e.
\begin{equation}
    {\bf s_{i}} = \sum_{j=1}^{m} {\bf \bar{x}_{j}} {c_{jv}},
    \label{NMF_simulated}
\end{equation}
where ${\bf s_i}$ is the $i^{\rm th}$ simulated vector in the matrix ${\bf S}$ of dimension $n~\times~v$, and {\bf $\bar{x}_{j}$} is a random sampling of the NMF features in ${\bf X}$ relative to the $j^{\rm th}$ NMF component.
This is the main concept on which this work is based. In what follows, we show how we decompose a line profile, ${\bf q_{i}}$, into its low-dimensional representation, ${\bf r_{i}}$, and then use the resulting NMF decomposition to create a set of synthetic spectra, ${\bf s_i}$.

\subsubsection{Implementation and tests of the NMF reconstruction}\label{ssubsec:quantitative_assessment}

We apply the formalism set above to our library of absorption line profiles, with which we compute the low-dimensional representation needed for profile generation. Key to this process is to determine how well the reconstructed values fit the observed ones. It is also important to quantitatively assess how reliable the new synthetic data are with respect to the observed spectra. These considerations set a twofold testing process to quantify the ability of the NMF in i) reconstructing the profiles, and ii) producing a new set of synthetic data. For the first task, we use the residual variance, $\sigma^2$, defined as the sum of the squares of the difference between the input profile, ${\bf q_{i}}$, and its reconstructed counterpart, ${\bf r_{i}}$. We also test the model accuracy by carrying out KS tests of the $\mathrm{\Delta\, V_{90}}$ distributions as traced by the input and reconstructed data. For the second task, we again use the KS test, setting as a requirement that the synthetic profiles must be characterised by a $\mathrm{\Delta\, V_{90}}$ distribution that is statistically consistent with that of the original profiles. These tests define our Key Performance Indicators (KPIs).

When applying this method to our library, we noticed that the variety of profiles in our sample, which is described by the large range in $\mathrm{\Delta\, V_{90}}$, affected the goodness of the NMF fitting. Running NMF on the entire sample resulted in a model with a high degree of complexity (high number of NMF components, or equivalently a high dimensional NMF space). This model ended up producing synthetic data not always similar to the observed profile shapes, thus failing to reproduce the input $\mathrm{\Delta\, V_{90}}$ distribution. To obviate this issue, we designed an algorithm that applies the NMF on subsets of profiles in smaller bins of $\mathrm{\Delta\, V_{90}}$, where the bins are selected adaptively by optimizing the two KPIs defined above.

We now describe the step-by-step procedure followed in designing this algorithm.
\begin{enumerate}
     \item {\it Definition of $\mathrm{\Delta V_{90}}$ bins} We define bins of $\mathrm{\Delta\, V_{90}}$ of increasing size varying as $\mathrm{{s}}_k = \mathrm{b_{edge}} + k\times20\, \mathrm{km\, s^{-1}}$, where $\mathrm{b_{edge}}$ represents the lower bound edge of the considered $\mathrm{\Delta V_{90}}$ bins and $k$ in the range $1 \le k \le 4$. We select the sub-sample of profiles, satisfying the condition $\mathrm{\Delta V_{90}}_{i} \le \mathrm{s}_k$ and on this, we perform multiple runs of NMF fitting, each with an increasing number of NMF components, $m$, specifically $2 \le m \le 30$. 

    \item {\it NMF fits} For each $\mathrm{\Delta V_{90}}$ bin, the NMF analysis returns the feature vectors ${\bf x_{j}}$ and their coefficient vectors in $\bf{C}$. The reconstructed profiles, ${\bf r_{i}}$, are then computed following Equation~\ref{NMF_spectrum}. Mock profiles are created by randomly sampling latent feature components from $\bf X$ (i.e., from each column, see Equation \ref{NMF_simulated}) to create the new latent feature matrix. As the number of artificial profiles created in this manner cannot exceed the size of the input data sample in that specific $\mathrm{\Delta\, V_{90}}$ bin, we carry out 100 different realizations of the simulation, each time sampling 66\% of the size of the input data.
    
    \item {\it KPIs analysis} The reconstructed and simulated samples are analysed in terms of $\mathrm{\Delta V_{90}}$ distributions and finally statistically compared with the distribution of $\mathrm{\Delta V_{90}}$ values of the input data sample using the $p$-values returned by the KS test. The performance of the reconstruction process is additionally tested by computing the mean residual variance between the input and reconstructed profiles, namely $<\sigma_i^2> = \frac{\sum_{i=1}^{n}{({\bf q_{i} - r_{i}})^2}}{n}$, with $n$ the number of input data. 
    
    \item {\it NMF model selection} To determine the optimal NMF model, we consider all the NMF representations that simultaneously satisfy the condition $p\mathrm{-value} > 0.1$ in both the data-$vs$-reconstructed and data-$vs$-simulated KS tests and for these we compare their $\sigma^2$ distributions as a function of the number of NMF components, $m$. A reasonable expectation is that $\sigma^2$ decreases in value as $m$ increases. However, an over-estimation of $m$ would include noise in the simulated profiles. As a solution, we consider the relative $<\sigma^2>$ improvement and select the optimal NMF model as the one for which we first measure an improvement larger than 40 percent. 
    In case the $p$-value condition is satisfied in multiple $\mathrm{\Delta V_{90}}$ bins, the optimal NMF model in each bin is selected as above, and finally the optimal bin size is chosen as the one in which the $<\sigma^2>$ value is the lowest. At the end of this step, the bin edge value, $\mathrm{b_{edge}}$, is updated to be the upper bound of the current step, and the process is repeated till the entire sample of data is analysed.
    
    \item {\it Problematic $\mathrm{\Delta V_{90}}$ bins} The procedure outlined above also identifies $\mathrm{\Delta V_{90}}$ bins in which the condition $p\mathrm{-value} > 0.1$ is never met. These are bins of $\mathrm{\Delta V_{90}}$ values in the ranges $80-100,\, 100-120, 120-140\,  \mathrm{km\, s^{-1}}$.
    As one would expect, the most dominant factor that can cause the NMF to fail is the diversity in the complexity of the input data, which we can parameterise with the number of Voigt components. Thus, we further divided the data falling in such problematic bins into low- and high-number of Voigt components subsets. We stress that low-/high-number of Voigt components does not imply low/high $\mathrm{\Delta V_{90}}$ values as it can be seen from Figure \ref{Sample} (bottom panel), where profiles with similar velocity widths are characterised by significantly different numbers of Voigt components. Thus, we split the two categories such that each contains roughly 50 percent of the total profiles in the bin. Once the division is done, we repeat the procedure outlined earlier to find the optimal NMF decomposition. 

\end{enumerate}



\begin{figure*}
	\includegraphics[width=14. cm]{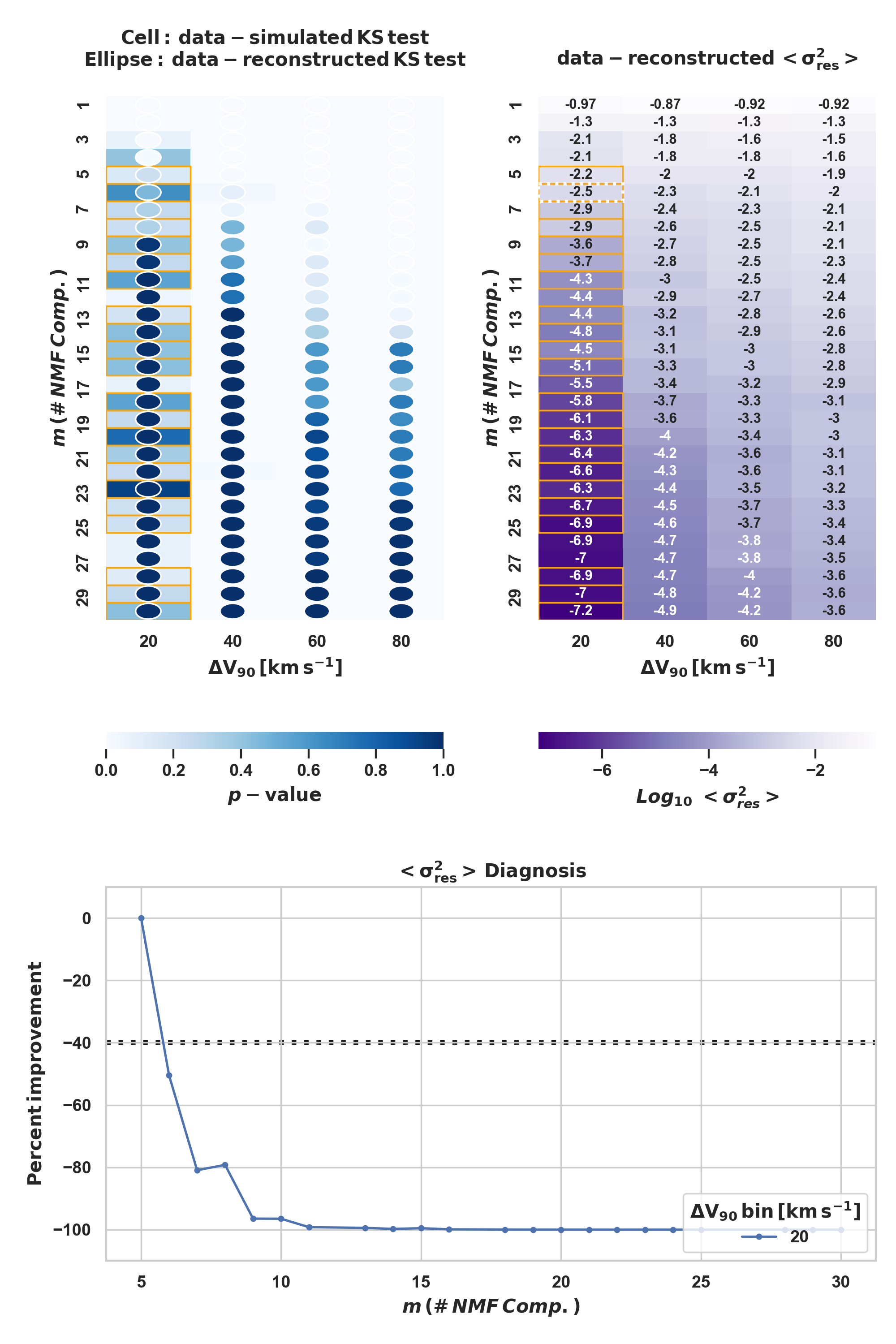}
    \caption{Examples of the performance metrics (KPIs) in the NMF reconstruction and simulation process for profiles characterised by small values of velocity widths ($\mathrm{\Delta V_{90}\le 80\, km\, s^{-1}} $). {\it Top-left:} Data-$vs$-reconstructed (grid cells) and data-$vs$-simulated (ellipses) KS tests of the $\mathrm{\Delta V_{90}}$ distributions as a function of the number of NMF components, $m$, and $\mathrm{\Delta V_{90}}$ bins of the input data. The colour code follows the KS test $p-$value statistics, with $p-{\rm value} > 0.1$ the threshold we use for statistical significance. Orange framed regions are where $p-{\rm value} > 0.1$ for both data-$vs$-reconstructed and data-$vs$- simulated distributions. {\it Top-right:} Same as the top-left panel, with the map coloured by the mean residual variance ($\log{< \sigma ^2 >}$) between the input and reconstructed profiles. The white dotted frame identifies the selected NMF model. {\it Bottom:} Relative $< \sigma ^2 >$ improvement between NMF models with an increasing number of NMF components (blue dots with line) for the $\mathrm{\Delta V_{90}}$ bin where the conditions $p-{\rm value} > 0.1$ is satisfied as given in the legend. The gray horizontal line identifies the 40 percent threshold in $<\mathrm{\sigma^{2}_{res}}>$ improvement we use to avoid NMF over-fitting.}
    \label{fig:working_bins_low}
\end{figure*}

\begin{figure*}
	\includegraphics[width=14. cm
	]{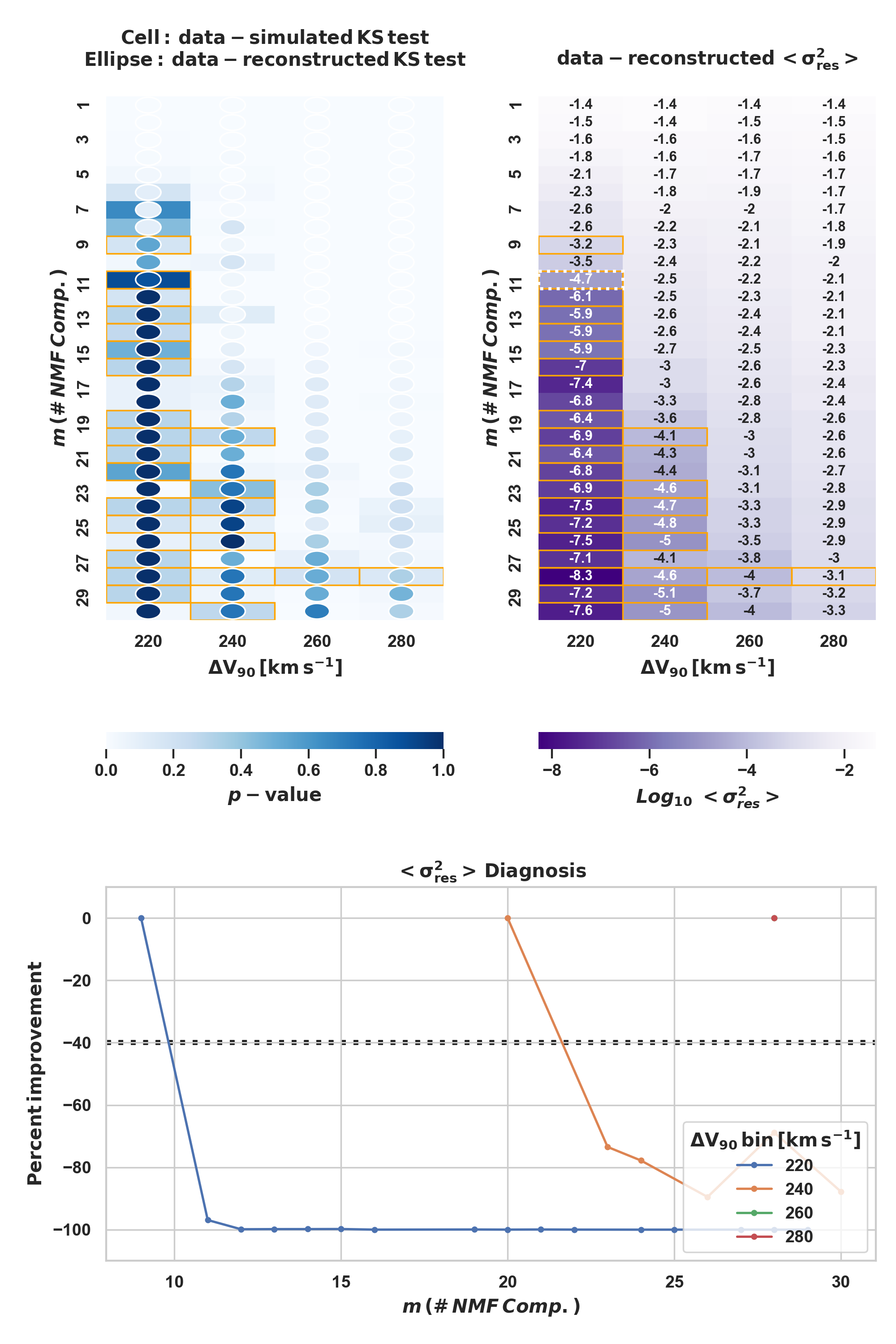}
    \caption{Same as Figure \ref{fig:working_bins_low} but for data characterised by larger velocity widths, $200 < \mathrm{\Delta V_{90} \le 280\, km\, s^{-1}}$. In this example, the $p-$value condition is satisfied in multiple $\mathrm{\Delta V_{90}}$ bins (as given in the legend). Dots with lines show the relative $< \sigma ^2 >$ improvement between NMF models with increasing number of NMF components. Relative to the last two  $\mathrm{\Delta V_{90}}$ bins only one NMF configuration satisfies our criteria, i.e. $m =27$, resulting in a single value in the $< \sigma_{\mathrm{res}} ^2 >$ Diagnosis plot and the two points, red and green, overlapping}.  
    \label{fig:working_bins}
\end{figure*}

\begin{figure*}
	\includegraphics[width=14 cm]{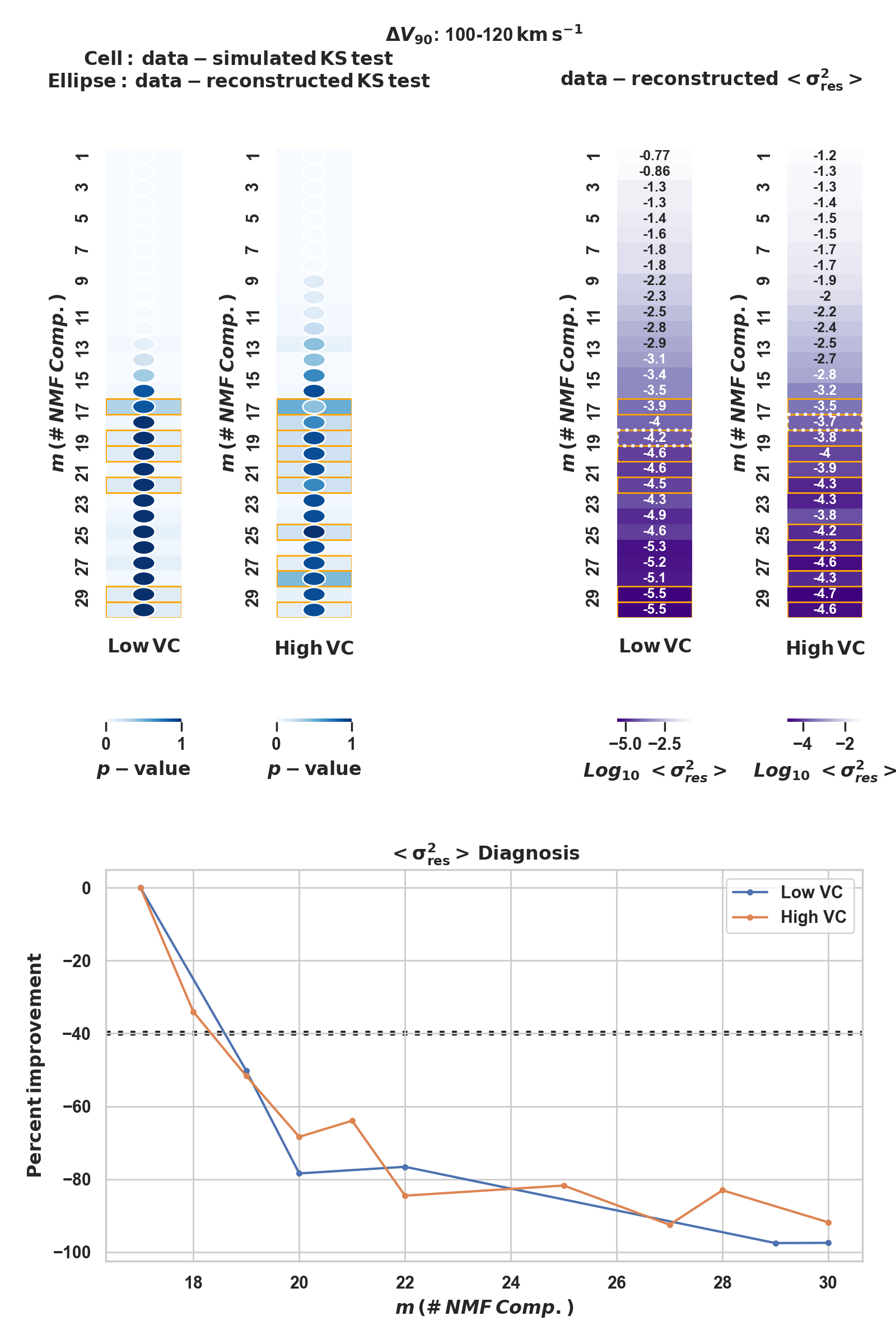}
    \caption{Same as Figures \ref{fig:working_bins_low}-\ref{fig:working_bins}, but for input data falling in the problematic bin with $\mathrm{100< \Delta V_{90}\le 120\, km\, s^{-1}}$. KS tests (top-left) and residual variance estimations (top-right) are carried out on the low- and high-number of Voigt components. }
    \label{fig:problematic_bins}
\end{figure*}

Examples of the procedure described above are shown in Figure~\ref{fig:working_bins_low} and Figure~\ref{fig:working_bins} (the KPI analysis run over the entire sample of data is provided in the online material) where, our KPIs for the NMF fitting and modelling are presented for a subsample of data falling in increasing size of $\mathrm{\Delta V_{90}}$ bins used in the iterative process that determines the final bin to select and, within this, the optimal NMF model.
The profiles characterised by low values of $\mathrm{\Delta V_{90}}$ are always preferred to be grouped together in the smallest bin size of $\mathrm{20\, km~s^{-1}}$. For example, the NMF decomposition on all the profiles with $\mathrm{ \Delta V_{90}} \le 80\, \mathrm{km\, s^{-1}}$  would succeed in the reconstruction step (i.e., $p-\mathrm{values} > 0.1$), but it would fail in generating synthetic profiles with the targeted $\mathrm{\Delta V_{90}}$ distribution (i.e., at least one NMF model for which $p-\mathrm{values} > 0.1$ is present). On the other hand, running the procedure on the sample of data for which $\mathrm{ \Delta V_{90}}~\le~20\, \mathrm{km\, s^{-1}}$ identifies multiple NMF models (shown as black framed in the top-left panel of Figure \ref{fig:working_bins_low}) that simultaneously satisfy the condition $p\mathrm{-value} > 0.1$ in both the data versus reconstructed and data versus simulated KS tests. Thus, the algorithm selects the NMF model with a $<\sigma^2>$ improvement closest to 40 percent (bottom panel in Figure~\ref{fig:working_bins_low}), i.e. the model with $m=6$ NMF components (dotted-white frame in the top-right panel of Figure~\ref{fig:working_bins_low}).

At larger $\mathrm{ \Delta V_{90}}$ values ($\mathrm{>180\, km~s^{-1} }$ ), the NMF fitting is less sensitive to the bin size. A clear case is shown in Figure \ref{fig:working_bins}, where both the reconstructed and simulated profiles are statistically consistent in following the same $\mathrm{ \Delta V_{90}}$ distributions as the one traced by the input data in bins of size $\mathrm{20,\, 40,\, 60,\, and\, 80\, km~s^{-1}}$. As described above, by analysing the variation of the relative $<\sigma^2>$ improvement we are able to avoid NMF over-fitting, and finally the optimal bin size is chosen as the one in which the difference between the input and the reconstructed profiles is the lowest (lowest value of $<\sigma^2>$ as shown in 'Data - Reconstruction residual variance' plot). In our example (Figure \ref{fig:working_bins}), the optimal NMF fit is obtained for profiles with $200 < \mathrm{ \Delta V_{90}}/\mathrm{km~s^{-1}} \le 220$ with $m= 11$ NMF components.

Figure \ref{fig:problematic_bins} shows the NMF analysis relative to the profiles for which the analysis based on a simple division in bins of $\mathrm{ \Delta V_{90}}$ is not possible due to the complexity of their shapes. These are profiles with a spread in velocities mostly falling within the range $100~<~\mathrm{ \Delta V_{90}}/\mathrm{km~s^{-1}}~<~140$. For these profiles we take advantage of the information we have on the number of Voigt components used for their decomposition (see Section \ref{Voigt_fitting} for more details) and the NMF fitting is run separately on two different samples, namely the low- and high-Voigt components (i.e. high-VC) samples, defined such that each contains roughly 50 percent of the total profiles.

\begin{figure*}
\includegraphics[trim={0cm 0cm 0cm 0cm}, clip, width=\textwidth] {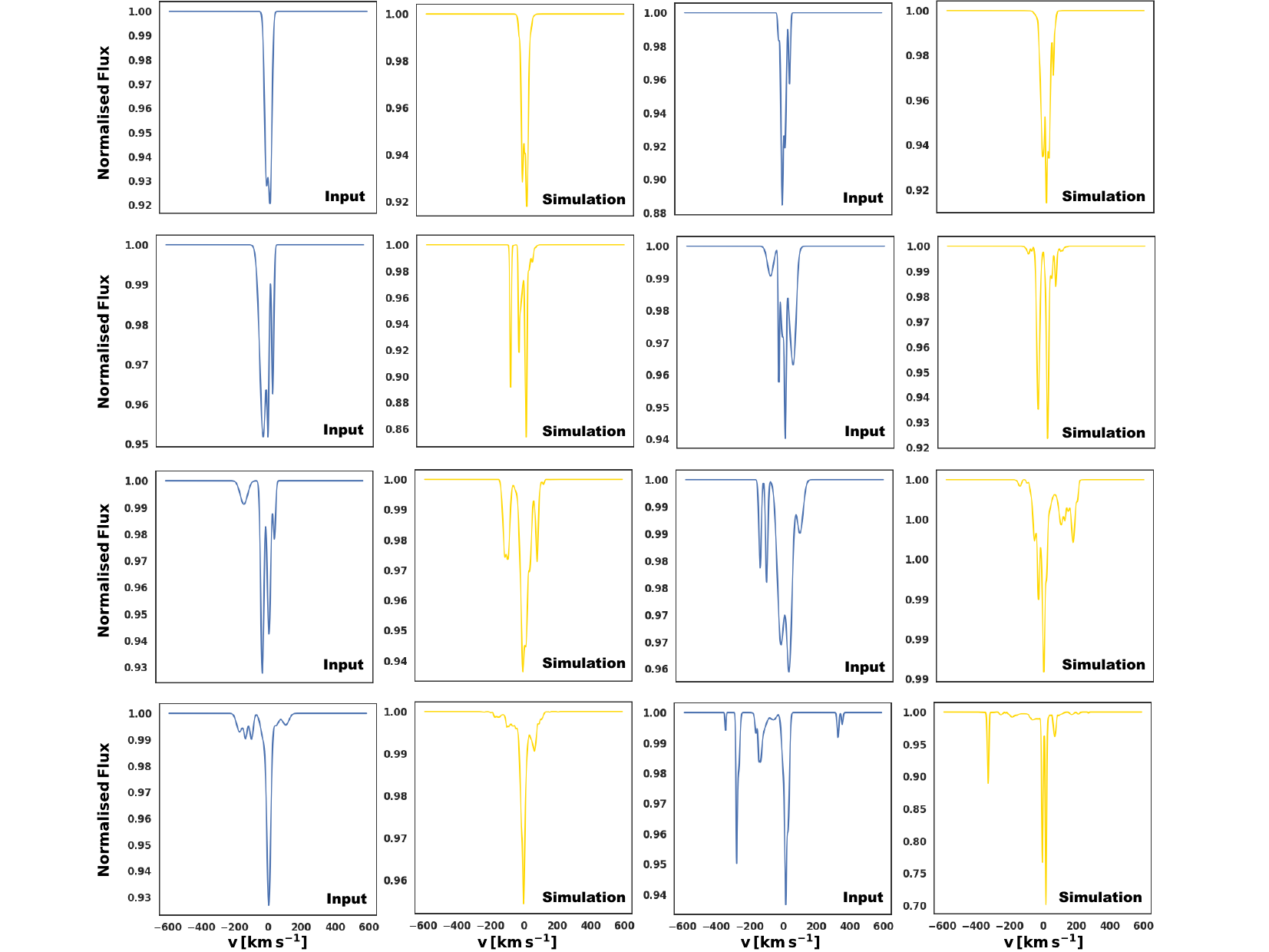}
    \caption{Comparison between input (blue) and simulated (gold) velocity profiles for a fixed column density of $\mathrm{\log(N/cm^{-2})} = 12$.}
    \label{fig:simualted_profiles}
\end{figure*}

\begin{figure}
	\includegraphics[width=\columnwidth]{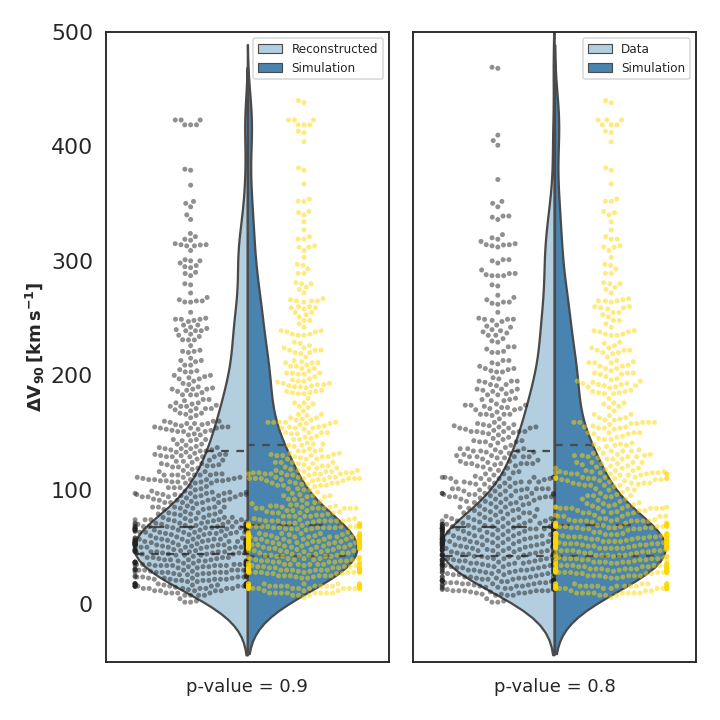}
    \caption{{\it Left panel:} Violin plots comparing the probability density of the $\mathrm{ \Delta V_{90}}$ distributions as traced by the reconstructed (light blue) and simulated (dark blue) profiles, where the central dashed line is the median and the dotted lines are the first and third quartiles. The entire distribution is shown as a swarm plot (dots). {\it Right panel:} Same as the left panel, but where the comparison is carried out between the input and simulated profiles. The $p-{\rm values}$ scores show that the simulated profiles are characterised by values statistically close to those traced by the reconstructed and input data. }
    \label{fig:Final_dv90_violin}
\end{figure}

\begin{figure}
\includegraphics[trim={19cm 0cm 3cm 0cm}, clip, width=\columnwidth] {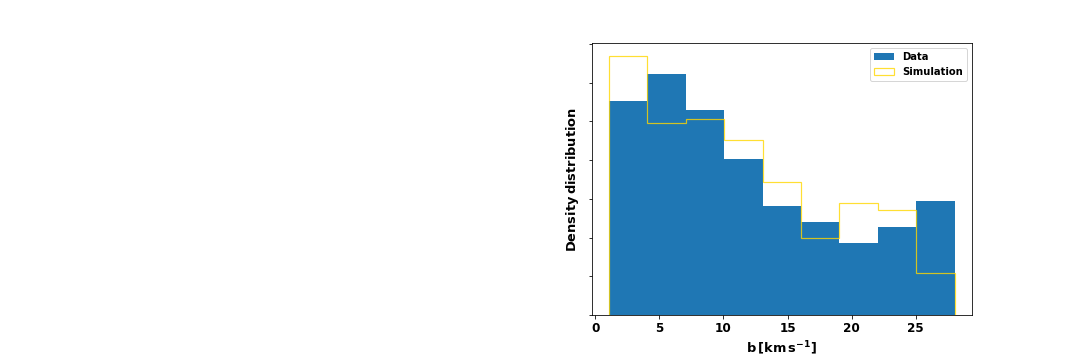}
\includegraphics[trim={2cm 0cm 19cm 0cm}, clip, width=\columnwidth] {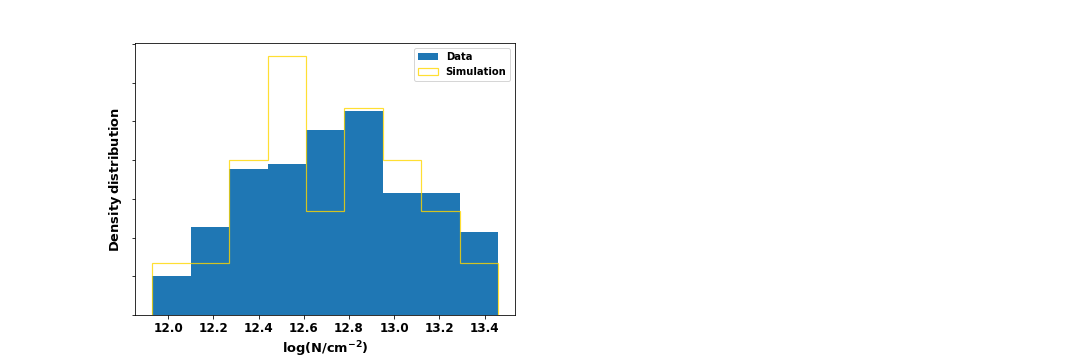}
    \caption{\textit{Top}: Normalised distributions of $b-$parameter values for observed (blue) and simulated (gold) velocity profiles. The simulations trace a family of high-resolution \ion{C}{IV} profiles, with $S/N =500$. \textit{Bottom}: Same as the top panel, however, this time the distributions are relative to column densities values. Here, the simulations trace high-resolution \ion{C}{IV} profiles, with  $15~<~S/N~<~30$, and $13 \le \mathrm{\log(N/cm^{-2})}\le 13.5$. $p-$value scores larger than 0.9 for both distributions show that the synthetic profiles have representative $b-$parameters and column densities values.}
    \label{fig:Voigt_component_test}
\end{figure}

To validate the procedure, we use the NMF algorithm described above to generate artificial profiles, and reproduce the input data in size, resolution (1 $\mathrm{km\, s^{-1}}$), and quality (infinite $S/N$). The goal is to compare synthetic and input data qualitatively and quantitatively by comparing their optical depths and $\Delta V_{90}$ distributions, respectively. As we have shown in Section \ref{ssubsec:quantitative_assessment} this is achieved by randomly sampling the NMF coefficients from their respective distributions of PDFs and finally using them as a new set of projections onto the NMF axes. The results of our simulations are presented in Figure \ref{fig:simualted_profiles}, where we show the comparison between synthetic (gold) and observed (blue) profiles, in terms of their optical depth distribution after having re-sampled them at a resolution of $8\, \mathrm{km~s^{-1}}$ (from left to right and top to bottom the profiles are characterised by increasing $\Delta V_{90}$ values). 
Our main result is that the diversity of the line profiles is well reproduced in the simulated sample despite the complexity of the input information. We evaluate the accuracy of our model by carrying out KS tests between the $\mathrm{\Delta V_{90}}$ distributions of the input  $vs$  simulated and reconstructed profiles, as shown in Figure \ref{fig:Final_dv90_violin}. We find that the simulated profiles have a $\mathrm{\Delta V_{90}}$ distribution statistically close to the one traced by the reconstructed ($p-{\rm value} = 0.9$) and input ($p-{\rm value} = 0.8$) samples.
Finally, in Figure \ref{fig:Voigt_component_test} we show that the individual components of the synthetic profiles have representative $b-$parameters and column densities when compared to the distributions traced by the observed absorbers ($p-$values $> 0.9$ for both samples). The evaluation for the $b-$parameters is carried out by running MC$-$ALF on 100 simulated \ion{C}{IV} profiles that are characterized by a resolution of  $8\, \mathrm{km\, s^{-1}}$, ideal $S/N=500$ values, and total column densities that follow the same distribution we measure for our input sample (see Figure \ref{Sample_dist} bottom panel). The check on the column density values, instead, is run on a smaller sample of 20 profiles, simulated with a realistic noise component, such that $15 < S/N < 30$ and moderately strong, i.e. with total column densities in the range $13 \le \mathrm{\log(N/cm^{-2})}\le 13.5$. Such a choice is for the results to be the least affected by uncertainties related to profile fitting. Indeed, for lower column densities it may become difficult to match single input-$vs$-retrieved Voigt components, while for higher values, single components may reach saturation and the column densities may no longer be estimated with a few percent accuracy (see Figure \ref{vc_testing1} central panel). The comparison is then carried out with a sample of real \ion{C}{IV} profiles with similar characteristics.

\subsection{The NMF-PM python package}

To enable the use of this tool by the community, we inserted the NMF feature matrix, {\bf X}, and coefficient matrix, {\bf C} in a python module, which we dubbed the NMF$-$PM (NMF Profile Maker). To run NMF$-$PM, the user will have to specify the number of simulated profiles to obtain in output via the parameter \texttt{nsim}, the ions to simulate, together with their rest-frame wavelengths and column density values passed via the parameters \texttt{ion}, \texttt{trans\_wl}, and \texttt{ion\_logN},  respectively. As discussed in Section \ref{statistics_library}, when considering different families of ions, i.e. moderate- and low-ion families, the absorbers may be characterised by different $\mathrm{ \Delta V_{90}}$ distributions. Thus one feature of the NMF$-$PM is to allow the user to specify which class of ion they are simulating via the parameter \texttt{ion\_family}. This can be set to '\textit{moderate}' or '\textit{low}' for the simulated profiles to follow a $\mathrm{ \Delta V_{90}}$ distribution as the one we measure for our samples of moderate- and low-ions (see Figure~\ref{Sample_dist}, top panel), or the user can feed their own $\mathrm{ \Delta V_{90}}$ PDF. NMF$-$PM also allows for the creation of ion doublets (e.g. \ion{C}{IV}, \ion{Mg}{II}) by setting to 'True' \texttt{doublets} and by providing a value for the \texttt{dbl\_fratio} and \texttt{dbl\_dvel} parameters for a given oscillator strength ratio and velocity shift (in $\mathrm{km\, s^{-1}}$) for the second line.

\begin{figure*}
    \centering
    \includegraphics[trim={0.25cm 0.1cm 0cm 0cm}, clip,width=\textwidth]{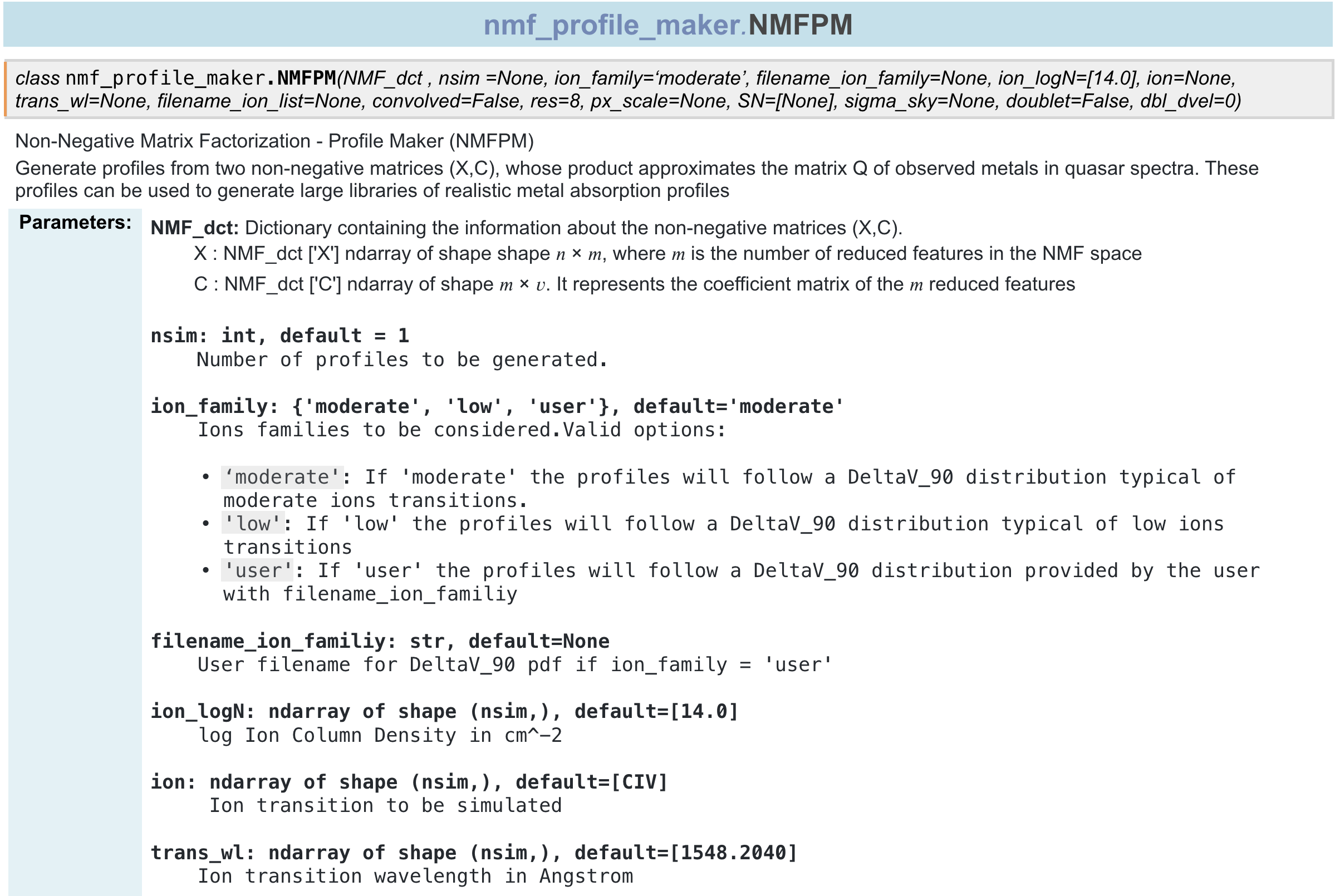}\\
    \vspace{-0.1cm}
    \includegraphics[trim={0.cm 3.8cm 0.3cm 0.1cm}, clip,width=\textwidth]{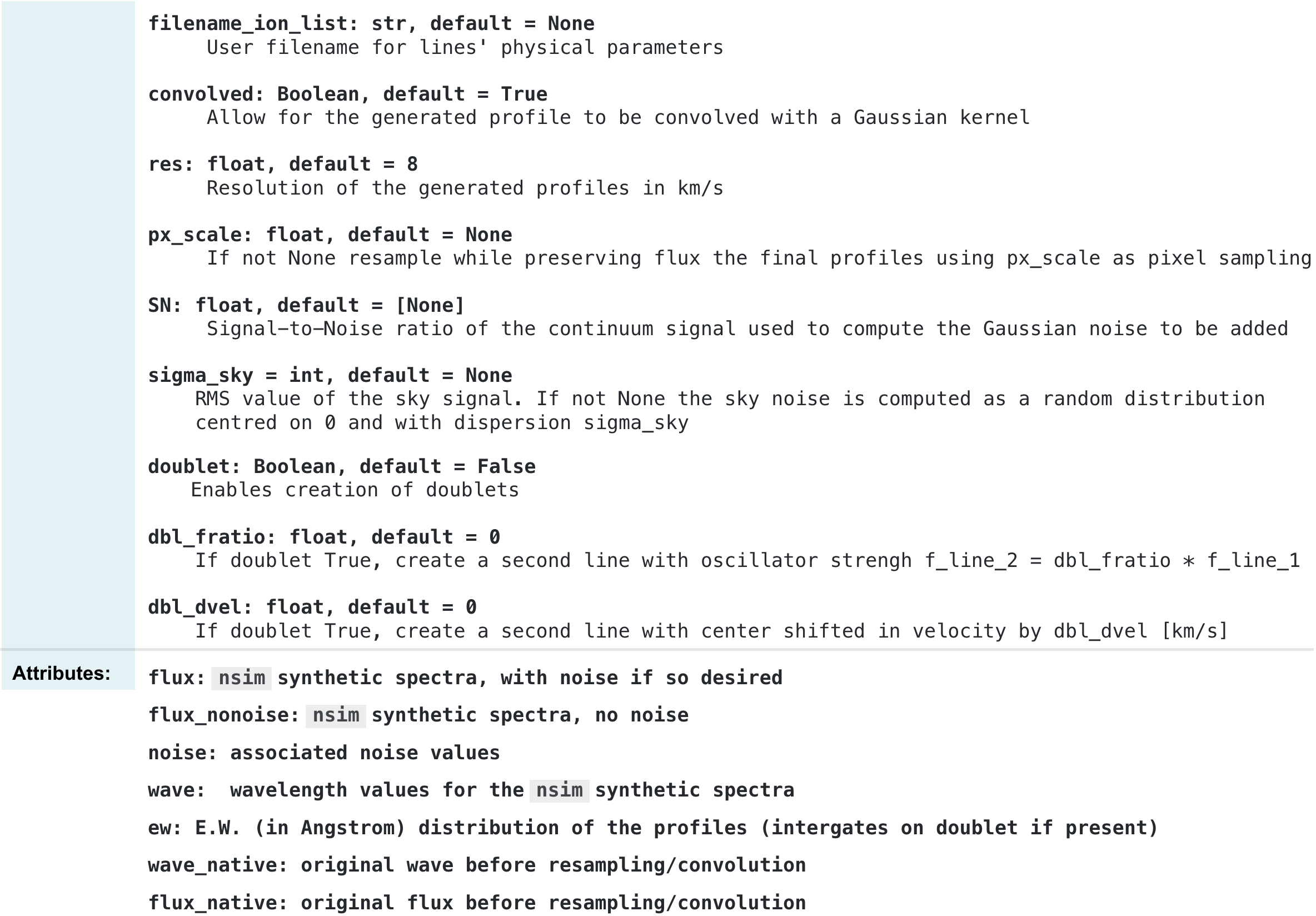}
    \caption{NMF$-$PM python class with parameters and attributes.}
    \label{NMF$-$PM_class}
\end{figure*}

With this configuration, NMF$-$PM simulates absorber profiles characterised by 1 $\mathrm{km\, s^{-1}}$ resolution and with no noise. However, the user can further: 1) convolve the profiles with a Gaussian kernel switching to 'True' the \texttt{convolved} parameter and consequently providing the resolution (FWHM) via \texttt{res}; 2) add a random Gaussian noise component by providing a value for the desired $S/N$ (per pixel) via \texttt{SNR}\footnote{This represents the $S/N$ ratio (per pixel) with respect to the continuum of the background source. To add a noise component relative to the sky signal the parameter \texttt{sigma\_sky} can be used.}, and 3) carry out a profile re-sampling providing a value for the \texttt{px\_scale} parameter. The re-sampling is implemented such that it conserves overall the integrated flux.
 Via its attribute, NMF$-$PM will return the convolved and re-sampled synthetic metal profiles with noise, and the associated noise and wavelength arrays. It will also return the original flux and wavelength arrays at a resolution of 1 $\mathrm{km\, s^{-1}}$ not convolved nor re-sampled. The NMF$-$PM class with its parameters and attributes is shown in Figure~\ref{NMF$-$PM_class}.

NMF$-$PM has been optimised to efficiently generate synthetic metal profiles: even performing the convolution step, the addition of Gaussian noise, and pixel re-sampling, the NMF$-$PM method runs in a matter of minutes on a single core computer to generate a library of $10^{5}$ objects.

\section{Simulated profiles in large surveys}
\label{NMF_results}

The automated tools we have described in this work open to the opportunity of testing our capability of detecting and analysing absorption features in spectra of different data quality in large surveys. To showcase the capabilities and further test the performance of our code for these applications, we run a library of  $10^{6}$ synthetic metal profiles mimicking \ion{C}{IV} absorbers using the moderate velocity distribution (see the red histogram in the top panel of Figure \ref{Sample_dist}). Profiles are generated in a flat distribution of column density in the interval $10^{13}-10^{15.5}~\rm cm^{-2}$, while the $S/N$ is uniformly distributed in the interval 2.5-15. We mock a WEAVE-like survey \citep{jin22} by setting the pixel scale to $16~\rm km~s^{-1}$ and the resolution to $60~\rm km~s^{-1}$.

\begin{figure*}
    \centering
    \includegraphics[width=0.49\textwidth]{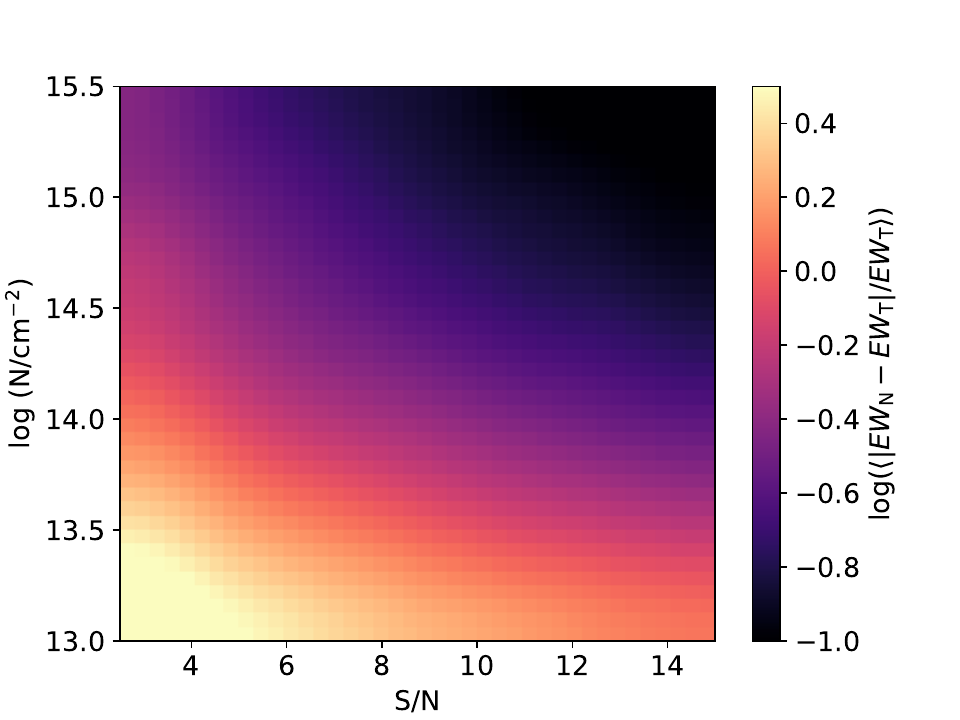}
    \includegraphics[width=0.49\textwidth]{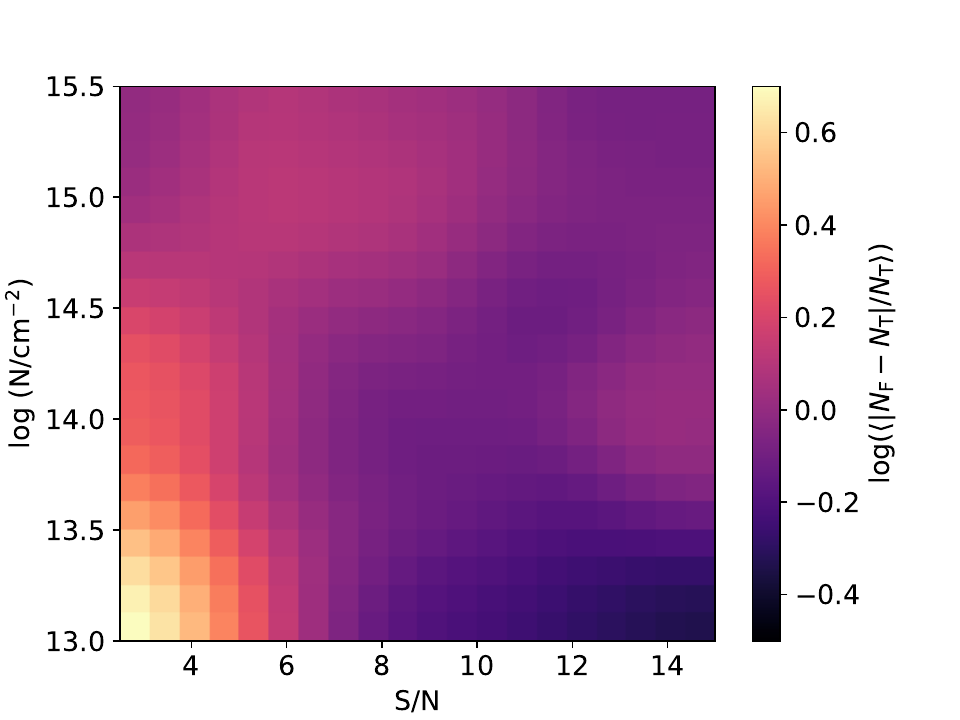}
    \caption{Test of detection and analysis of \ion{C}{IV} absorption features of different quality in WEAVE-like spectra. {\it Left Panel}:  Mean $\mathrm{EW}$ relative errors (on a logarithmic scale) as a function of $S/N$ and input column density values for $10^{6}$ test profiles. For clarity reasons, the metric has been smoothed with a $2\times2$ Gaussian kernel. {\it Right Panel}: Same as the left panel, however, this time the mean relative error statistics is shown for the column densities and for a subsample of $10^4$ profiles. Given the lower number statistics, the smoothing is carried out on a $3\times3$ kernel window. Values less or equal than zero identify the regions in the $\mathrm{\log(N/cm^{-2})}\, vs\, S/N$ plane of high precision in the fitted values. }
    \label{fig:CIV_WEAVE_test}
\end{figure*}

Figure \ref{fig:CIV_WEAVE_test} (left panel) shows the measured equivalent width ($\mathrm{EW}$) for the stronger line of the doublet measured on noisy profiles in comparison with the intrinsic value derived from the noise-free simulated profiles. To better capture the intrinsic scatter in the distribution rather than a possible bias, we compute the average of the absolute discrepancy, normalized by the true value. Values less than one in this metric identify $\mathrm{EW}$ values retrieved with high precision. These plots reveal the expected trend of increasing precision in the measurement as both column density and $S/N$ increase. At moderate $S/N$, or $\lesssim 4$, \ion{C}{IV} can be measured reliably only for column density $\gtrsim 10^{14}~\rm cm^{-2}$. 

\begin{figure}
    \centering
    \includegraphics[width=\columnwidth]{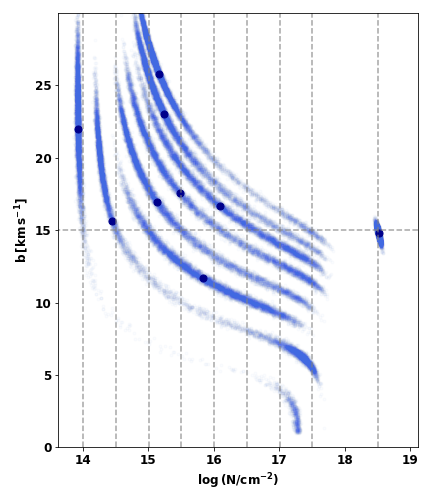}
    \caption{Same as Fig. \ref{test_sat}, however this time the $b\, vs\,  \mathrm{log{(N/ cm^{-2})}}$ plot is shown for saturated profiles at a resolution of $60\, \mathrm{km s^{-1}}$. The broader range of fitted parameters is an indication of hidden saturation at play, resulting in a larger relative error for the column density estimates ($<~\delta^{\mathrm{\log{N}}}> = 0.05~$).}
    \label{test_sat_WEAVE}
\end{figure}

Next, we proceed and fit a subsample of $10^4$ profiles with MC$-$ALF to study the accuracy in retrieving the column density using the same metric we used for the previous test. The results are shown in Figure \ref{fig:CIV_WEAVE_test} (right panel): excluding the bins characterised by both $3\le S/N \le 5$ and $\mathrm{\log(N/cm^{-2})} = 13.5$, for which the quality of the data prevent the fit to run correctly, MC$-$ALF can retrieve the input information at all $S/N$ and column density values considered. 
 In particular, fits of mildly saturated and unsaturated profiles with $\mathrm{\log(N/cm^{-2})} \le 14.5$  are less sensitive to variations in $S/N$ for $S/N > 7$. For $S/N \le 7$, the accuracy decreases, with the lowest values ($\sim 40\%$) measured for $S/N \sim 3$. For heavily saturated lines (in our example for $\mathrm{\log(N/cm^{-2})} > 14.5$), MC$-$ALF fits have larger uncertainties ($\sim 20\%$ for $S/N < 10$), although the accuracy increases as a function of the $S/N$. The effect of saturation on low-resolution spectra is further analysed by repeating the test presented in Figure \ref{test_sat} for \ion{C}{II} profiles this time convolved with a FWHM of $\sim 60\, \mathrm{km\, s^{-1}}$. The results (Figure \ref{test_sat_WEAVE}) show that, by exploring the full underlying posterior, MC$-$ALF is able to recover the degeneracy between the $b$-parameter and the column density estimates due to hidden saturation resulting in broad and degenerate posterior distributions. For saturated profiles, in the regime when the damping wings are not yet significant (in this example for $\mathrm{\log(N/cm^{-2})}< 18$), the full posterior PDF should be used for an accurate propagation of uncertainties.

\section{Summary and Conclusions}
\label{conclusions}

In this work, we present two new tools for studying and modelling metal absorption lines in the circumgalactic medium: MC$-$ALF to automatically reconstruct
the physical parameters of the absorbers and NMF$-$PM to generate synthetic
but realistic-looking line profiles following a given distribution of
desired line width.

The observational data we used for developing, training, and testing our codes come from a compilation of spectroscopical campaigns, which collected high resolution, high $S/N$ spectra of 42 quasar fields at redshifts $1.2 \le z \le 4.5$ (Section \ref{data}). These surveys identified a family of $\sim 1000$ moderate- and low-ion absorbers along the quasars' line of sight.
 By considering only unsatureted profiles, we selected a sample of 650 absorbers, with redshifts ranging from $z =0.9 - 4.2 $ and column densities in the range $11.2 \le \mathrm{\log(N/cm^{-2})} \le 16.3$. These represent our library of absorption line systems, which gathers a large variety of profiles in terms of shape and line widths.

Our tools rely on advanced numerical techniques.
MC$-$ALF uses a Bayesian approach to absorption line fitting, which, with minimal human intervention, can decompose metal lines into individual Voigt components providing the posterior distributions of the line parameters such as the column density, the Doppler parameter, and the redshift. Moreover, as the likelihood space is sampled via the nested sampling algorithm {\sc PolyChord}, MC$-$ALF is highly efficient in discriminating among competing models for profiles of different complexity (typically related to the instrument resolution and data $S/N$). Quality assurance tests on simulated UVES-like profiles demonstrate that MC$-$ALF is able to recover the input information with small relative errors: for the $b$-parameters, column densities, and redshifts distributions we find mean relative errors of $<\delta> = 0.03,0.002,\, \mathrm{and}\, 0.62\times 10^{-6}$, respectively. 

We next showed that Non Negative Matrix factorization (NMF) methods offer a straightforward statistical framework for physically relevant predictions of non-negative, continuous signals after the data have been properly standardised (Section \ref{data_standardization}). Moreover, as outliers can significantly impact NMF, we build a statistical framework to select the most appropriate bin in $\mathrm{\Delta V_{90}}$ to perform the fitting. The results are evaluated in terms of residual variance, $\sigma^2$, of the difference between the input profile and its reconstructed counterpart and Kolmogorov-Smirnov (KS) tests among the $\mathrm{\Delta\, V_{90}}$ distributions as traced by the input, reconstructed, and simulated data. We then inserted the NMF feature matrix, {\bf X}, and coefficient matrix, {\bf C} in the NMF$-$PM python module with which the user can simulate $10^{6}$ metal profiles following a given distribution of desired line width in approximately 10 minutes  (on a one core machine). 

Upcoming wide-field surveys, like DESI, 4MOST, and WEAVE, are taking the challenge of observing an unprecedented sample (around a million) of quasar spectra to detail the properties and the evolution of the galaxies' CGM across the Universe. This work aims at contributing to the scientific effort of simulating, testing the detection, and calibrating the observations of metal absorbers in large quasar surveys. In particular, we have shown that our tools will make it possible to reliably simulate, identify and characterise both weak and strong metal absorption lines even in a low-resolution regime. This will, in turn, enable the study of a large sample of lower and higher column density and/or higher redshift systems to resolve small- and large-scale CGM effects and their relation with the surrounding larger-scale environment \citep[e.g.][]{lofthouse20,lofthouse23,dutta20} and to target regions in the Universe at a key epoch for galaxy formation and evolution. 
On the basis of making our modelling easily accessible to the large astronomical community, we make publicly available MC$-$ALF and NMF$-$PM  that will allow any user to produce a library of synthetic profiles and analyse them with a simple click of a key. MC$-$ALF and NMF$-$PM are available on the github pages provided in the Data Availability Section.

\section*{Acknowledgements}
This project has received funding from the European Research Council (ERC) under the European Union's Horizon 2020 research and innovation programme (grant agreement No 757535) and by Fondazione Cariplo, grant No 2018-2329.

\section*{Data Availability}
 The authors provide the KPI analysis run over the entire sample of data (see Section \ref{ssubsec:quantitative_assessment}) as online material, while the two tools MC$-$ALF and NMF$-$PM are publicly available at the github pages \texttt{github.com/matteofox/MC-ALF/} and \texttt{github.com/alongobardi/NMFPM/}, respectively. 
Finally, the library of $\sim 10^{6}$ synthetic WEAVE-like profiles will be shared upon request to the corresponding author.




\bibliographystyle{mnras}
\bibliography{Biblio} 

\bsp	
\label{lastpage}
\end{document}